# From Displacement to Angle: Diamond-Based 3D Rotation Sensing for High-Precision Cellular Force Measurement

*Linjie Ma[1][#], Bicong Wang[2][#], Tai Nam Yip[1], Yong Hou[1]\*, Yuan Lin[2]\*, and Zhiqin Chu[1,3]\**

1. Department of Electrical and Electronic Engineering, The University of Hong Kong, Hong Kong SAR
2. Department of Mechanical Engineering, The University of Hong Kong, Hong Kong SAR
3. School of Biomedical Sciences, The University of Hong Kong, Hong Kong SAR

\*Corresponding authors:
Dr. Yong Hou, Email: houyong@eee.hku.hk
Prof. Dr. Yuan Lin, Email: ylin@hku.hk
Prof. Dr. Zhiqin Chu, E-mail: zqchu@eee.hku.hk

[#] These authors made equal contributions to this work.

## Abstract

Cellular traction forces are conventionally measured by tracking the displacement of beads or micropillars, an approach fundamentally limited by optical diffraction and the classical Euler-Bernoulli beam assumption, which is accurate only when the traction-induced deformation is relatively small while the aspect ratio of micropillars is large. Here we introduce an alternative approach: quantifying force through direct measurement of rotational angle, in addition of displacement of the micropillar, using fluorescent nanodiamonds as embedded 3D orientation markers. Specifically, by integrating optically detected magnetic resonance (ODMR) with laser polarization modulation (LPM), we determine the complete three-dimensional orientation of nanodiamonds attached to PDMS micropillars with sub-degree precision (~0.5°). This angle-based measurement framework bypasses the resolution constraints of displacement tracking and remains valid for stocky beams or when large deformations occur. Finite-element simulations demonstrate that our method reduces force estimation errors by at least 10% compared to conventional displacement-based approaches. Moreover, we successfully capture multidimensional pillar deformations—including bending and twisting—that are inaccessible to conventional displacement-only method. Taken together, our work establishes diamond-based angular force microscopy as a high-precision platform for mechanobiology.

# 1. Introduction

Cells generate mechanical forces on their surrounding matrix to regulate crucial cellular activities, such as spreading, migration, division, and differentiation[1-6]. Accurate measurement of cell–substrate forces is therefore essential for understanding how cells sense and respond to their mechanical microenvironment[7-9] and has motivated the development of a variety of traction force measurement platforms [10]. Conventional traction force measurement methods, including bead tracking in gels [11-13] and micropillar arrays [14], infer cellular forces from the displacement of markers (i.e., beads and micropillars) and are limited in precision by optical diffraction and assumptions of geometrical/material linearity. For example, localizing a fluorescent object with ~300 nm lateral resolution imposes a fundamental limit on how finely forces can be mapped, and while super-resolution techniques can push beyond this barrier, they come at the cost of speed, complexity, and throughput [15]. Moreover, the standard linear relationship between displacement and force assumes that deformations are small enough for the geometry to remain largely unperturbed, a condition often violated in practice.

Rather than pursuing ever higher precision within the same translational measurement framework, we ask whether a different mechanical observable can provide a more informative and robust readout of cellular forces. In displacement-based methods, mechanical sensors are effectively treated as point-like probes, whose motion is reduced to pure translation, leaving rotational degrees of freedom unmeasured even though they are intrinsically coupled to how forces are transmitted [16]. For example, in standard pillar-based traction force microscopy, each PDMS post is modeled as an independent cantilever whose lateral force is inferred solely from tip displacement via a linear spring calibration $P = 3EI\delta/L^3$, derived from small-deflection Euler–Bernoulli beam theory[17]. Yet any bending of a pillar necessarily produces not only a lateral deflection of the tip but also a finite tilt of the pillar axis in the vertical (x–z) plane. Conventional analyses use only the translational component $\delta$ and discard this tilting part. In addition, micropillars fabricated from PDMS, a material commonly used in mechanobiology, exhibit low bending stiffness. Consequently, cellular forces may induce relatively large deformations at the pillar top, thereby compromising the applicability of the linear force–displacement relationship. These limitations motivate an angle-based readout in which the bending-induced rotation angle replaces translational displacement for mechanical measurement. This approach becomes particularly advantageous under large pillar bending, where conventional displacement-based models become unreliable.

To realize such an angle-based readout, we need a method that can report the full three-dimensional orientation of nanoscale markers on pillar tops with sub-degree precision under diffraction-limited imaging. Recently, diamond-based quantum sensors have emerged as powerful tools for tracking multi-dimensional motions in biological systems [18-19]. The negatively charged nitrogen–vacancy center (NV$^-$ center, hereafter abbreviated as NV center) [20] in diamond acts as a built-in sensor whose spin resonance is highly sensitive to the relative orientation of external magnetic field and NV center axis [21-23], enabling off-axis rotation detection via optically detected magnetic resonance (ODMR, angular resolution∽0.5° [24]), while its fluorescence is polarization-dependent [25-26], allowing in-plane rotation tracking via laser polarization modulation (LPM, angular resolution∽2–3° [16]). Using these methods, NV centers in diamond have been successfully used as orientation markers to measure nanoscale substrate deformation in force fields [16, 24]. However, initial demonstrations primarily measured in-plane rotation of nanodiamonds in cell traction fields, providing only a two-dimensional angular projection and preventing direct force quantification from rotation alone. Extending this capability to complete three-dimensional orientation sensing, which is essential for resolving multidimensional forces and establishing robust angle-to-force calibration, remains challenging.

Here, we develop a hybrid ODMR–LPM method that determines the complete three-dimensional orientation of fluorescent nanodiamonds attached to PDMS micropillars with sub-degree precision (≈0.5° for out-of-plane rotation and 3° for in plane rotation). By directly measuring the bending-induced rotation angle, we establish an angle-based force calibration framework that circumvents both the diffraction limit and the linearity constraints of conventional displacement tracking. An analytical treatment based on the exact differential equation governing the deflection of micropillars (irrespective of their aspect ratios) provides us a relationship between the measured angle and the applied force. Finite-element simulations demonstrate that this approach reduces force estimation errors by at least 10 % compared to displacement-based methods, with the largest improvements occurring for low-aspect-ratio pillars where geometric nonlinearity is most pronounced. We further validate the 3D rotational tracking capability by using bulk diamond calibration standards, confirming an angular precision of ~0.5° for out-of-plane rotations. Finally, we apply this platform to living cells and successfully resolve their traction fields.

## 2. Results and Discussion
### 2.1. Theoretical validation of angle as a better mechanical readout in pillar-based traction force microscope

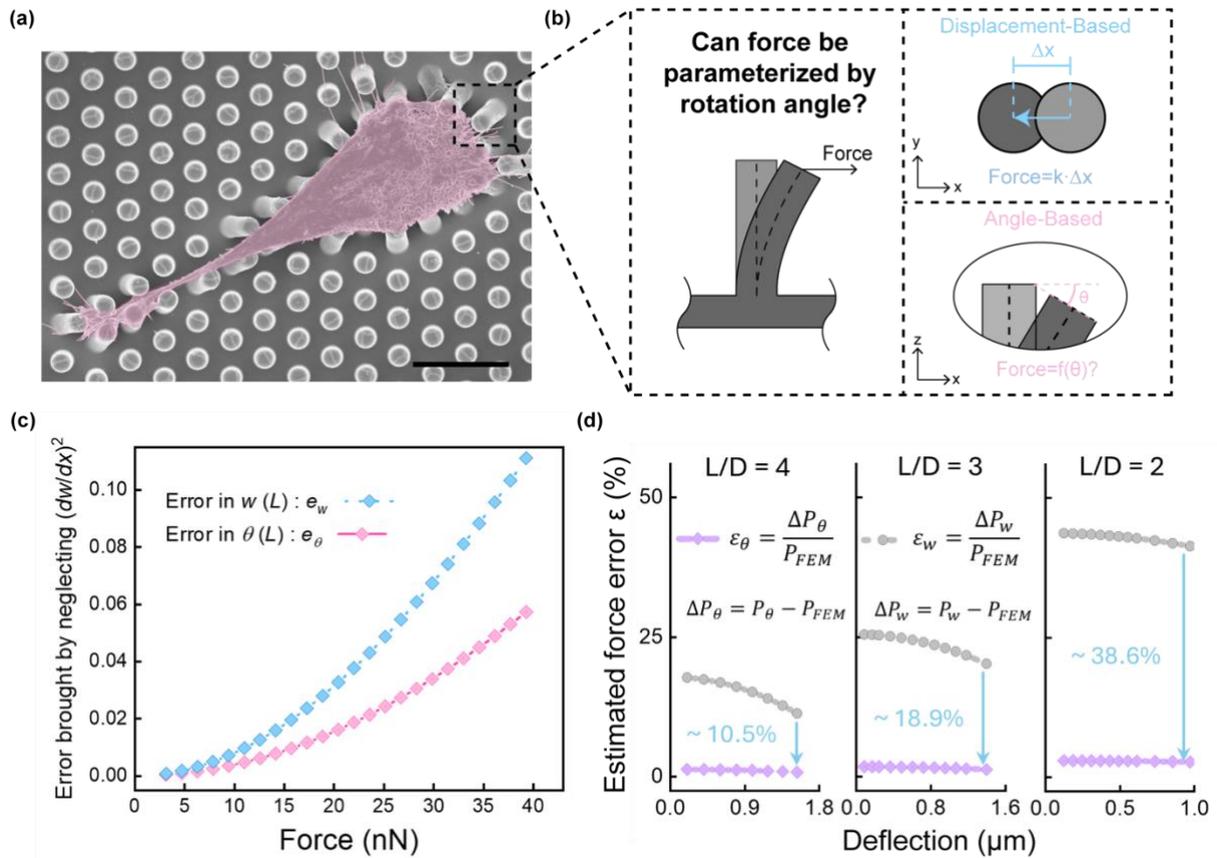

**Figure 1.** Rotation angle as an alternative high-accuracy mechanical readout for cellular force measurement. a) SEM image of bent PDMS pillars caused by cell traction force. The scale bar is 10 μm. b) Comparison of the traditional displacement-based method and the newly proposed angle-based method. c) The errors brought by $1+(dw/dx)^2 \approx 1$ within the framework of Euler-Bernoulli beam theory. d) The errors in estimated forces with pillar length $L$ of 6μm and diameter $D$ of 1.5μm, 2μm and 3μm. $P_w$ and $P_\theta$ represent the force estimated from pillar deflection and rotation angle, respectively, and $P_{FEM}$ represents the force calculated from finite element simulations.

The classical displacement-based force estimation relies on the Euler-Bernoulli beam theory, which assumes the deformations are relatively small so the cross-section of the beams remains a plane after deflection. For a cantilever beam (**Figure S1**), the relationship between the concentrated force $P$ at the free end and deformation ($\theta$ and $w$ for rotational angle and deflection, respectively) can be described by a differential equation [27]:

$$\frac{d^2w}{dx^2}\frac{1}{\left(1+\left(\frac{dw}{dx}\right)^2\right)^{\frac{3}{2}}} = \frac{d\theta}{dx}\frac{1}{\sqrt{1+\left(\frac{dw}{dx}\right)^2}} = \frac{P(L-x)}{EI}, \tag{1}$$

where $E$ represents the Young's modulus, $I$ denotes the inertia moment of the beam cross-section. Based on the small deformation assumption, the terms of $(dw/dx)^2$ in the denominators of **Equation 1** can be disregarded, leading to:

$$w_{EB}(L) = \frac{PL^3}{3EI}, \quad \theta_{EB}(L) = \frac{PL^2}{2EI}. \tag{2}$$

The deviations incurred at the free end of the beam due to neglecting the $(dw/dx)^2$ term can be estimated by (see Supplementary Materials for details):

$$e_\theta = \left|\frac{\theta_{EB}-\theta_{IF}}{\theta_{IF}}\right| \approx \frac{1}{1+\frac{24E^2I^2}{P^2L^4}}, \quad e_w = \left|\frac{w_{EB}-w_{IF}}{w_{IF}}\right| \approx \frac{1}{1+\frac{35E^2I^2}{3P^2L^4}}. \tag{1}$$

which suggests that, within the scope of Euler-Bernoulli beam theory, the deviation in the linear force-displacement relationship is more pronounced than that in the force-angle relationship since $e_\theta < e_w$ even when the applied force is small (**Figure 1c**). Consequently, the angle can provide a more accurate mechanical readout than the displacement. Moreover, it is also assumed in small deformation that the arc coordinate $s$ (along the deformed beam) and the reference coordinate $x$ approximately equal to each other, something will no-longer be appropriate when beam deflection becomes large. Actually, it can be shown that, even under large deformation, the load $P$ acting on a micropillar can be expressed in terms of its rotation angle $\theta_L$ at the free end as:

$$F(k) - F(k,\psi_0) = \sqrt{\frac{P}{EI}}L, \quad \psi_0 = arcsin\frac{1}{\sqrt{2}k}, \quad k = \sqrt{\frac{sin\theta_L+1}{2}}, \tag{2}$$

where $F(k)$ and $F(k,\psi_0)$ respectively represent the complete and incomplete elliptic integral of the first kind (see Supplementary Materials for details [15]). In practice, the pillar arrays made of polydimethylsiloxane (PDMS) are monolithic with the substrate. Therefore, the measured rotation angle of the pillar top $\theta_M$ comprises contributions from bending of the beam $\theta_L$ and tilting of the substrate $\theta_T$. We here introduced a factor (see Supplementary Materials for details):

$$C_\theta = \frac{1}{\frac{4(1-\nu^2)D}{3\pi}\frac{1}{L}+1} \tag{3}$$

to eliminate the effect of tilting of pillar base. The corrected angle $\theta_L = C_\theta\,\theta_M$ can be directly substituted into **Equation 4** to quantify the lateral force applied on the pillar top. With finite element simulations, we demonstrated that the newly proposed angle-based force estimation method achieves exceptional accuracy, reducing the error of the displacement-based method by up to ~40% for micropillars with small aspect ratios (**Figure 1d**).

## 2.2. Integrating ODMR and LPM: Enabling 3D orientation tracking of fluorescent nanodiamonds

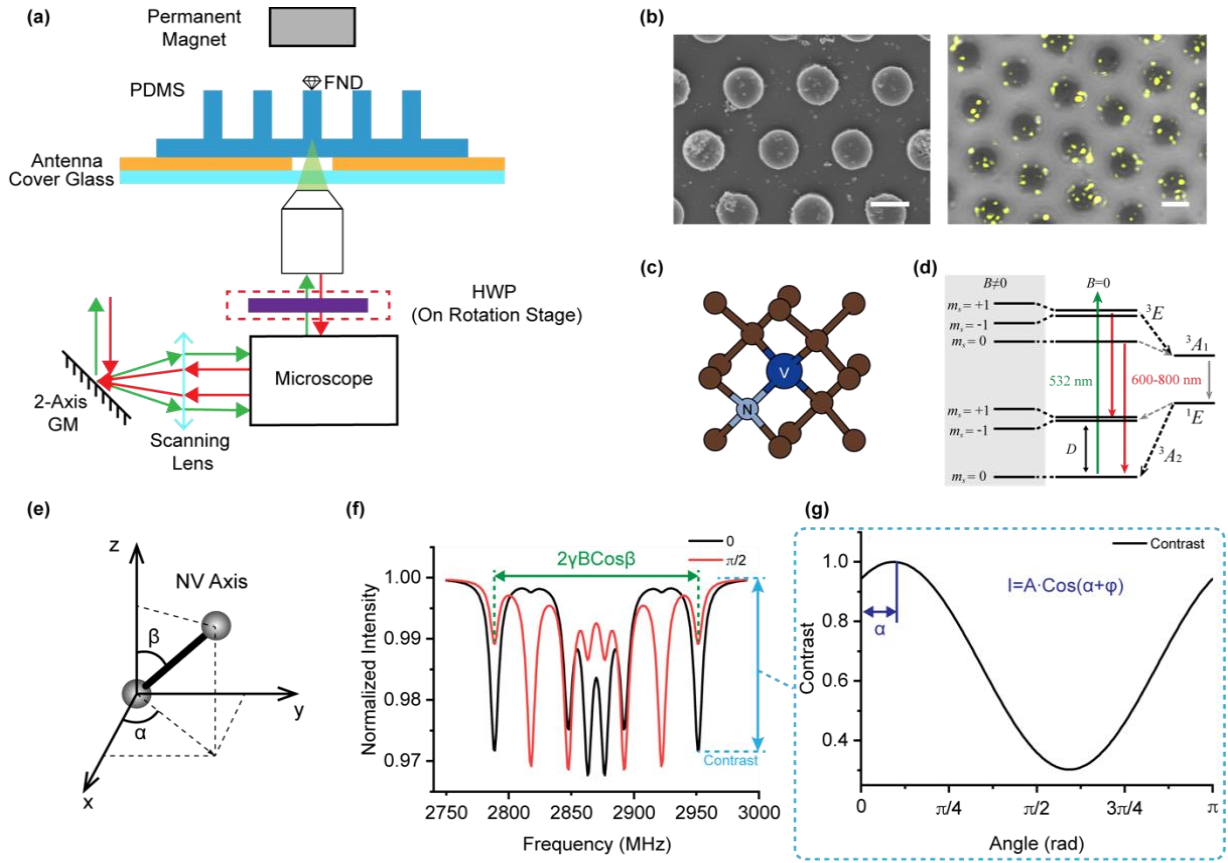

**Figure 2.** Quantum-enabled three-dimensional rotation measurement using ODMR-LPM hybrid method. a) Optical path design diagram. b) SEM image (left) and confocal image (right) of PDMS pillar array with nanodiamond. The scale bar is 2 μm. c) Illustration of NV center

structure. d) Energy level of NV center. e) Orientation of NV axis and corresponding angle. f) Simulated ODMR spectrum for NV centers under different laser polarizations. g) ODMR contrast response of laser polarization.

To translate the theoretical advantages of angle-based force measurement into practice, we use fluorescent nanodiamonds (FNDs) containing NV centers as three-dimensional rotation sensors. By combining the magnetic-field-dependent resonance of NV centers with their polarization-dependent optical response, we integrate optically detected magnetic resonance (ODMR) and laser polarization modulation (LPM) to overcome the limitations of each technique when used alone. As shown in **Figure 2a**, a home-built confocal microscope system combining polarized optical excitation, microwave modulation, and an external magnetic field was developed for ODMR–LPM angle measurements. FNDs were fixed on the tops of PDMS micropillars (**Figure 2b**), serving as orientation markers that convert pillar bending and torsion into measurable changes in crystal orientation of FND.

The NV center is a point defect in diamond consisting of a substitutional nitrogen atom adjacent to a vacancy (**Figure 2c**). Its spin-triplet ground state exhibits Zeeman splitting under an external magnetic field (**Figure 2d**), and the magnitude of this splitting depends on the angle between the NV axis and the magnetic-field direction. This enables the determination of the relative orientation between each NV axis and the field. However, rotation around the magnetic-field axis does not change this relative angle, making axial rotation inaccessible with a single magnetic field. Although multi-field ODMR approaches can resolve this ambiguity, they significantly increase experimental complexity.

All-optical polarization-based methods, such as LPM, determine NV orientation through polarization-dependent excitation or emission. It is sensitive to the in-plane projection of the NV axis but cannot fully resolve the three-dimensional orientation. Moreover, for nanodiamonds containing ensembles of NV centers with multiple orientations, the superposition of polarization responses results in a single projected vector. According to Euler's rotation theorem, complete determination of rigid-body rotation requires at least two non-collinear vectors; therefore, LPM alone cannot reconstruct the full rotation matrix.

By integrating the ODMR and LPM techniques, the limitations inherent to each individual method can be overcome. The orientation-dependent Zeeman splitting enables frequency-domain separation of NV centers with different crystallographic axes, while LPM provides the rotation information around the magnetic-field axis. To implement this strategy, a polarized laser, an external microwave, and a magnetic field are combined within a confocal microscope platform (Figure 2a). A half-wave plate mounted before the objective controls the excitation polarization. The microwave is fed to the sample through a co-planar waveguide deposited on the cover glass. The magnetic field is generated by a permanent magnet, and its direction is carefully aligned to be vertical (calibration is shown in the SI). In this way, the diamond orientation can be separated into two independent angles (**Figure 2e**): a vertical angle $\beta$, extracted from the ODMR peak separation (**Figure 2f**), and a horizontal angle $\alpha$, determined from the polarization-dependent contrast variation of the corresponding ODMR peaks (**Figure 2g**). The determination of $\alpha$ and $\beta$ is detailed in SI. Because NV centers with different orientations are separated in the frequency domain, their polarization responses can be individually analyzed. With 2 or more NV centers with different orientations, the rotation of the diamond particle can be fully determined.

**2.3. Validation of In-plane and Out-of-Plane rotation angle measurement**

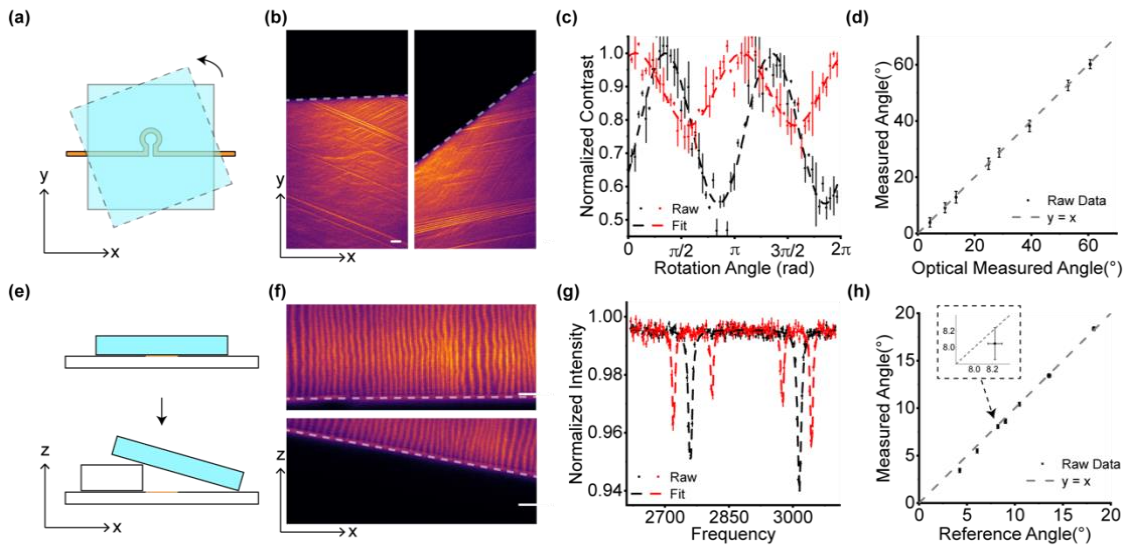

**Figure 3.** Experimental validation of the proposed ODMR-LPM hybrid method for 3D rotation measurement. a) Schematic of the in-plane rotation of the bulk diamond. b) Confocal image of the bulk diamond before (left) and after (right) the rotation. The edge of the diamond is labeled with a dashed line. c) Typical measurement result of the LPM Spectrum. The black point and line show the measured data and fitting result of the diamond before rotation. The red points and line show the measured data and fitting result of the diamond after rotation. d) Comparison of the measured in-plane rotation angle using the optical method and the LPM-ODMR method. e) Schematic of the out-of-plane rotation of the bulk diamond. f) Confocal image of the bulk diamond before (left) and after (right) the rotation. The edge of the diamond is labeled with a dashed line. g) Typical measurement result of the ODMR Spectrum. The black points and line show the measured data and fitting result of the diamond before rotation. The red points and line show the measured data and fitting result of the diamond after rotation. h) Comparison of the measured out-of-plane rotation angle using the optical method and the LPM-ODMR method. The scale bars in (b) and (d) are 10 μm.

To validate the proposed rotation angle measurement method and evaluate its performance, we conducted both horizontal and vertical rotation experiments using a bulk diamond sample. As shown in **Figure 3a**, a [100]-cut bulk diamond was placed on a cover glass coated with a coplanar waveguide. The diamond was randomly rotated in the horizontal plane, and at each rotation angle, both confocal scanning and LPM-ODMR measurements were performed. The confocal image provides an optical reference for the LPM-ODMR measurement. As shown in **Figure 3b**, due to the presence of color centers in the bulk diamond, a clear boundary corresponding to the edge of the diamond can be observed in the confocal image. The horizontal rotation angle was determined by comparing the boundary in images acquired at different rotation angles. During the LPM-ODMR measurements, the ODMR spectrum showed almost no change, which can be explained by the method's mechanism: the relative angle between the NV centers and the applied magnetic field remains constant under in-plane rotation. In contrast, the phase shift in the LPM spectrum clearly reveals the horizontal rotation, as shown in **Figure 3c.** By comparing the optical reference angle with the rotation angle extracted from the LPM-ODMR method (**Figure 3d**), the horizontal rotation measurement demonstrates high accuracy. The vertical rotation measurement method was validated in a similar manner. As shown in **Figure 3e**, one edge of the bulk diamond remained in contact with the cover glass, while a thin quartz plate was inserted beneath the opposite side, causing the diamond to rotate by a certain angle in the vertical direction. The vertical rotation angle could be adjusted by varying the thickness of the inserted quartz plate. To determine the exact rotation angle, an XZ-plane scan was performed with the confocal microscope, allowing visualization of the cross-section of the bulk diamond, as shown in **Figure 3f**. The rotation angle obtained by comparing the boundaries in two images was treated as the optical reference, and the corresponding rotation angle was also extracted from the LPM-ODMR measurement. Since the relative orientation between the NV centers and the magnetic field changes during out-of-plane rotation, a significant shift in the ODMR spectrum was observed (**Figure 3g**). By comparing the optical reference with the rotation angle measured by the LPM-ODMR method (**Figure 3h**), the vertical rotation measurement shows high accuracy, with an error of approximately 0.5°, comparable to previously reported results from conventional ODMR methods.

## 2.4. Multidimensional force measurement in cells

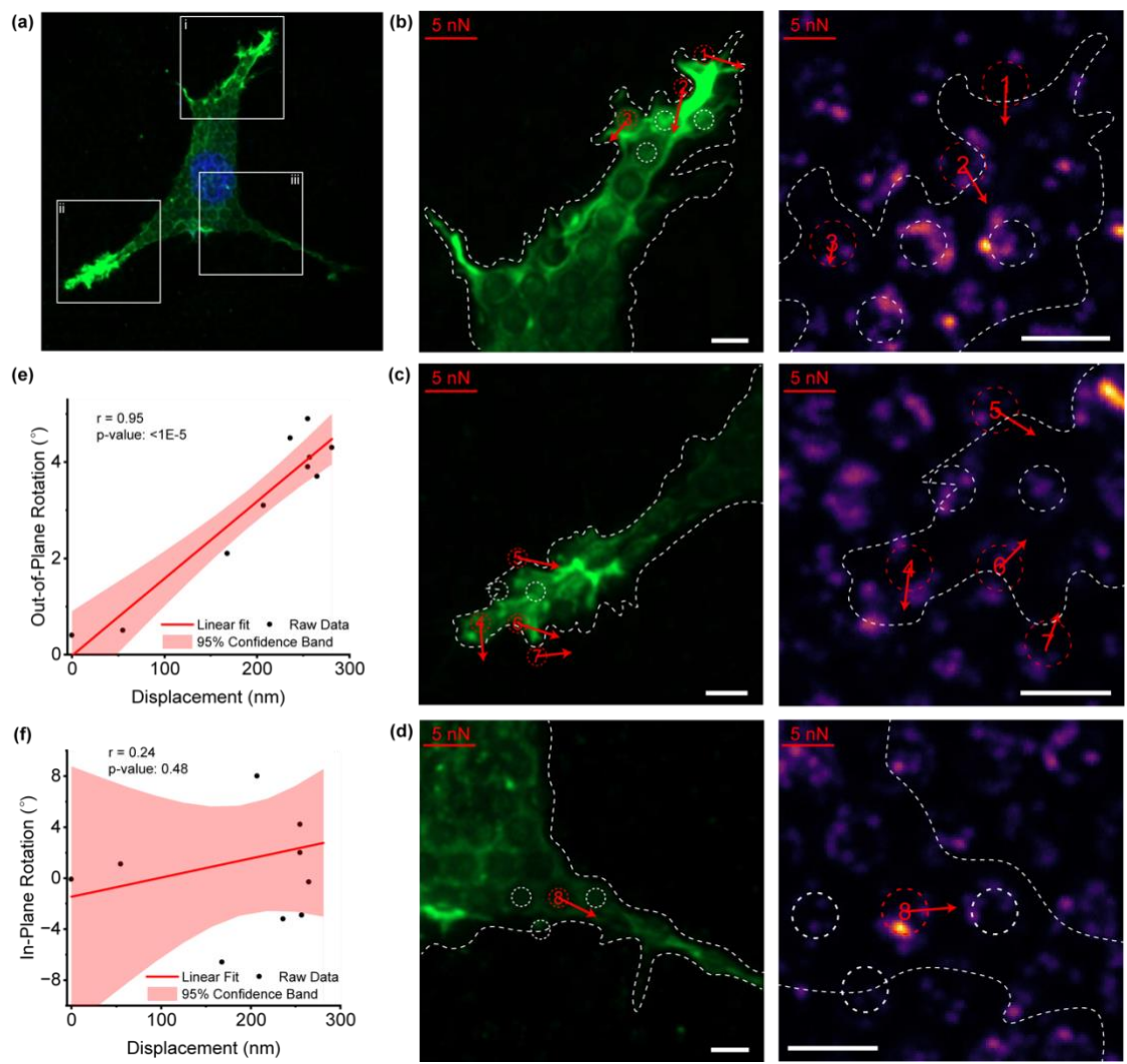

**Figure 4.** Multi-dimensional cellular force measurement using ODMR-LPM hybrid measurement method. a) Confocal image of the measured cell and three measured areas. b-d) Cell force measurement using the displacement method (left panels) and the ODMR-LPM hybrid method (right panels). The scale bars in all figures are 4 μm. e) Correlation analysis of out-of-plane rotation and displacement. f) Correlation analysis of in-plane rotation and displacement.

|  | In-Plane Rotation | Out-of-Plane Rotation | Force (Angle Method) | Displacement | Force (Displacement Method) | Relative Difference* |
|---|---|---|---|---|---|---|
| Pillar 1 | 4.2° | 3.9° | 4.0 nN | 255 nm | 4.1 nN | -2.4% |
| Pillar 2 | 13.0° | 4.3° | 4.4 nN | 281 nm | 4.6 nN | -4.3% |
| Pillar 3 | −6.6° | 2.1° | 2.1 nN | 168 nm | 2.7 nN | -22% |
| Pillar 4 | −2.9° | 4.1° | 4.2 nN | 257 nm | 4.2 nN | -0% |
| Pillar 5 | −3.2° | 4.5° | 4.6 nN | 236 nm | 3.8 nN | +21% |
| Pillar 6 | −0.3° | 3.7° | 3.8 nN | 265 nm | 4.3 nN | -12% |
| Pillar 7 | 8.0° | 3.1° | 3.2 nN | 207 nm | 3.4 nN | -5.9% |
| Pillar 8 | 2.0° | 4.9° | 5.0 nN | 255 nm | 4.1 nN | +22% |
| Control 1 | 1.1° | 0.5° | | 55 nm | | |
| Control 2 | −0.1° | 0.4° | | 0 nm | | |

**Table 1**. Cell traction forces from angle and displacement methods. *The relative difference is calculated by the formula: $\frac{Force(Angle\ Method) - Force(Displacement\ Method)}{Force(Displacement\ Method)}$.

With the accuracy and precision of our 3D angular sensing platform validated, we next applied it to a relevant biological context: measuring the complex forces exerted by cells. We seeded NIH-3T3 fibroblasts onto nanodiamond-functionalized PDMS micropillar arrays. After 24-hour incubation, cells spread and generated contractile forces, inducing multidimensional pillar deformations. We subsequently fixed the cells, lysed them with proteinase K to release mechanical stress, and allowed the elastic pillars to return to their original positions. During this stress-relaxation process, we simultaneously measured out-of-plane bending angles and in-plane rotation angles of pillar tops using LPM-ODMR, and recorded nanodiamond displacements via confocal imaging for conventional displacement-based TFM analysis.

At polarized regions of spread cells−including areas 1, 2, and 3 where traction forces typically concentrate−we detected significant rotational signals. Specifically, out-of-plane rotations ranged from 2.1° to 4.9°, while in-plane rotations ranged from 2° to 13°. In contrast, non-adhesion control regions exhibited out-of-plane rotations below 0.5° and in-plane rotations below 1.1° (within measurement uncertainty), confirming the specificity of our mechanical readouts.

Correlation analysis revealed distinct relationships between rotational motions and pillar displacement (**Figure 4e and f**). Out-of-plane rotation exhibited a strong positive correlation with displacement, indicating that horizontal traction forces dominate pillar bending. Conversely, in-plane rotation showed no significant correlation with displacement, suggesting that torsional forces arise from more complex mechanical interactions. We hypothesize that in-plane rotation originates from cellular traction but is strongly modulated by the eccentricity of force application−theoretically, forces applied farther from the pillar central axis generate larger torsional moments. Consistently, pillars 1, 2, 5, and 7, where cellular adhesions localized near the upper edge of pillar tops, displayed more pronounced rotational motions due to their longer effective lever arms compared to pillars beneath the cell body.

More importantly, our multidimensional angular measurement system enables precise quantification of cellular traction forces on individual pillars. Angle-derived forces were consistently lower than displacement-based estimates (**Table 1**), a trend that aligns quantitatively with our theoretical predictions and finite-element simulations. This systematic discrepancy arises because displacement methods, constrained by optical resolution limitations, overestimate actual pillar deformation under identical imaging conditions.

In summary, our angular measurement platform successfully captures multidimensional mechanical signatures−bending (out-of-plane rotation) and torsion (in-plane rotation)−in cellular force fields, and establishes a rotation-based framework for accurate force quantification. This work provides a new paradigm for deciphering the complex mechanobiology of cell-matrix interactions.

## 3. Discussion

The central advance of this work is a conceptual shift in pillar-based traction force microscopy: from displacement as the primary mechanical observable to rotational angle. Conventional micropillar and gel-based traction force measurements infer cellular forces from lateral displacements, typically under a linear force-deflection assumption. This approach is fundamentally constrained by optical diffraction and image registration accuracy, and linear beam theory imposes requirements on the geometry and deformation of the pillar. In contrast, angular measurement exploits the orientation-dependent optical response of NV centers in nanodiamonds, transforming the spatial resolution problem into an angular resolution problem. In our implementation, we achieve ~0.5° angular precision, which is equivalent to resolving ~30 nm arc displacement at the top of a 6 μm pillar (L/D =3, 1.49 MPa), far exceeding what can be reliably resolved by conventional diffraction-limited imaging under similar conditions [28-29].

From a mechanical standpoint, our analysis clarifies why the angle a better observable. Within the classical Euler–Bernoulli theory framework, neglecting the $(dw/dx)^2$ term introduces a smaller relative error in the predicted rotation than in the predicted tip displacement, leading to systematically more accurate force estimates when angle is used as the calibration variable. Moreover, when large deformations occur and the beam axis is significantly curved, the load–angle relationship remains expressible through elliptic integrals, while the displacement-based linear calibration diverges. Finite-element simulations confirm that our angle-based method reduces force errors by at least 10% compared to displacement-based estimates. Thus, angle-based readout simultaneously relaxes diffraction-limited spatial resolution requirements and mitigates model errors associated with geometric nonlinearity, making it both conceptually and quantitatively advantageous under realistic cellular loading conditions.

Beyond improving force accuracy, angle readout fundamentally enriches the type of mechanical information accessible from micropillar arrays. The three-dimensional orientation of the pillar top encodes both out-of-plane bending and in-plane torsion, allowing us to distinguish between centrally applied forces and off-axis traction that generates torque. In our cell experiments, we observed in-plane rotations that did not correlate with tip displacement, particularly on pillars where focal adhesions were located near the edge of the pillar top. This observation is consistent with a simple lever-arm picture: forces applied farther from the pillar axis produce larger torsional moments for the same net traction. Such torsional signals are largely invisible to conventional displacement-only readouts, yet they may carry important information about how cells organize contractile actin bundles and focal adhesions to generate both forces and moments on their substrate. In future work, systematic mapping of bending and torsion across different cell types and perturbations could reveal whether torque generation is a regulated component of mechano-transduction—for example, in directional migration, matrix remodeling, or durotaxis.

At the same time, the present implementation of the ODMR–LPM platform has several technical limitations that motivate further development. First, the external magnetic field is provided by a fixed permanent magnet, which ensures excellent stability but limits field uniformity and alignment flexibility across the field of view. Spatial variations in field magnitude and direction can introduce systematic errors in the extracted NV orientation. Replacing the permanent magnet with a well-characterized electromagnet or Helmholtz-coil configuration would improve field homogeneity and allow dynamic control of the field direction, increasing robustness and enabling more flexible measurement protocols. A second limitation is temporal resolution: the combined ODMR–LPM readout, as implemented here, relies on sequential spectral and polarization scans, which constrains both imaging speed and throughput. For many biological questions, such as fast force fluctuations during protrusion–retraction cycles or dynamic remodeling of traction during migration, real-time or near-real-time 3D force readout will be desirable. Advances in widefield ODMR imaging, rapid microwave and polarization modulation, and parallel readout of different NV ensembles provide clear routes toward higher-speed, higher-throughput implementations suitable for tracking dynamic cellular force generation in living systems.

Taken together, our results establish rotational angle as a rigorous and experimentally accessible mechanical observable for micropillar-based traction force measurements and demonstrate that diamond-based 3D orientation sensing provides a practical route to implement this concept. By integrating NV-center nanodiamonds with PDMS micropillar arrays, we transform a conventional displacement sensor into a quantum-enhanced, rotation-angle-based cellular force probe capable of resolving both forces and torques at the single-pillar level. This rotation-based framework extends the capabilities of micropillar platforms from scalar force readout toward high-precision, multidimensional biomechanical sensing. With further technical refinement, especially in magnetic-field control and parallel, high-speed readout, angular traction microscopy offers a promising pathway for investigating complex, time-dependent mechanobiological processes and microscale material responses.

**4. Conclusion**

In conclusion, we introduce an angle-based framework for quantifying cell traction forces using PDMS micropillar arrays. Theoretical analysis shows that rotational angle estimation produces smaller errors than displacement-based approaches and remains valid for pillars with different aspect ratios. This provides a solid mechanical foundation for replacing conventional pillar-top displacement tracking with rotation-angle-based force reconstruction. By integrating diamond-based orientation sensing with a hybrid ODMR–LPM measurement method, we directly measure the rotational angle at the pillar top. This approach transforms the conventional PDMS micropillar array into a quantum-enhanced rotation-angle-based cellular force sensor. Moreover, the proposed framework extends the capabilities of micropillar platforms toward high-precision, multidimensional biomechanical sensing. With further technical refinement, it provides a promising pathway for investigating complex biological and material systems at the microscale.

**5. Methods**

*Sample Preparation*: An omega-shaped coplanar microwave antenna was fabricated directly on a glass coverslip by thermal evaporation of a Ti/Au/Ti (titanium/gold/titanium) tri-layer [30]. The PDMS pillar array was fabricated using standard soft lithography techniques. Each pillar measured 6 μm in height and 2 μm in diameter. PDMS (Sylgard 184, Dow Corning) was mixed with its curing agent at a 10:1 ratio and poured onto 1 cm × 1 cm silicon molds. The mixture was degassed under vacuum for 30 min and subsequently covered with the glass coverslip contained coplanar microwave antenna. Samples were cured in an oven at 80 °C overnight to achieve a Young's modulus of $1.4 \pm 0.5$ MPa. The cured samples were peeled off the molds. The pillar arrays were cylindrical in shape, with a center-to-center spacing equal to twice the pillar diameter, thereby maintaining an ordered area density.

To ensure that the PDMS layer on the coverslip remained thinner than 50 μm, the amount of PDMS dispensed onto the silicon molds was controlled within 5–10 mg. This limitation was required to ensure that the total thickness did not exceed the working distance of the microscope objective.

*Nanodiamond Coating*: The surface of the PDMS pillar array was functionalized by immersion in 3-aminopropyl triethoxysilane (APTES) solution in ethanol (8%, v/v) for 10 min, followed by sequential rinsing with ethanol (99.8%) and deionized (DI) water. Fluorescent nanodiamonds (FNDs; 0.025 mg/mL in DI water) were dispersed using an ultrasonic water bath. The PDMS pillar array was incubated with the dispersed FND solution for 10 min, after which unbound FNDs were removed by rinsing with DI water.

*Confocal Measurement*: The confocal images in this study were acquired using two different setups. To locate fluorescent nanodiamonds (FNDs) for ODMR and LPM experiments, we used a custom-built confocal microscope (as described in the Supplementary Information) equipped with an air objective (UPLANXAPO40X, Olympus) with a numerical aperture of 0.95 NA 532 nm continuous-wave green laser provided excitation at a power of 5 μW.

For illustrative purposes, including high-resolution imaging of the cell and sample structure and three-dimensional optical sectioning, we used a commercial confocal microscope (LSM 980, Zeiss).

*ODMR Measurement*: The microwave required for the ODMR experiments was generated by a microwave source (SynthNV Pro, Windfreak Technologies), amplified using a high-power amplifier (ZHL-16W-43-S+, Mini-Circuits), and delivered to the sample through an omega-shaped coplanar waveguide patterned on the glass coverslip. The reference channel was used to monitor baseline fluorescence for normalization purposes, while the signal channel captured the microwave-modulated fluorescence response of the NV centers. The total integration time for each frequency point is 54 ms, and each loop is repeated for 10 times.

*LPM measurement*: In the LPM measurements, a mechanical rotation stage was used to rotate a half-wave plate (HWP), thereby modulating the polarization direction of the excitation laser. During the experiment, the rotation speed was set to 15 degrees per second. While the stage was rotating, the previously described microwave sequence was continuously fed into the sample, and both the reference fluorescence intensity and the signal intensity were simultaneously recorded in two separate photon counting channels. The rotation stage completed a full 360° cycle, and fluorescence intensity data were recorded at 3° intervals.

*Cell Culture*: NIH-3T3 cells were cultured in T-25 flasks. The complete growth medium consisted of DMEM (Gibco) supplemented with 10% (v/v) fetal bovine serum (FBS; Invitrogen), 2 mM L-glutamine (Invitrogen), and 100 U/mL penicillin-streptomycin (Invitrogen). Cells were maintained at 37 °C in a humidified incubator with 5% $CO_2$.

Prior to cell seeding, the nanodiamond sensor array was coated with human plasma fibronectin solution (50 μg/mL; Thermo) and incubated at 4 °C overnight. The array was then rinsed three times with phosphate-buffered saline (PBS) and sterilized under UV light for 20 min. Cells were trypsinized, resuspended in DMEM, and seeded onto the sensor array for 4–6 h.

*Fluorescence staining*: Cells cultured on the sensor array were fixed with 4% paraformaldehyde for 10 min at room temperature. Following fixation, cells were permeabilized with 0.25% Triton X-100 in PBS for 5 min at room temperature.

Nuclear staining was performed using Hoechst 33342 (Invitrogen; 1:700 dilution), and filamentous actin was labeled with Phalloidin (Abcam, iFluor 488, ab176753; 1:1000 dilution). Staining was carried out for 1 h at room temperature. Finally, cells were rinsed three times with PBS and imaged using a Zeiss LSM 980 confocal microscope.

**Acknowledgements**

Z.Q.C. acknowledges the financial support from the National Natural Science Foundation of China (NSFC) and the Research Grants Council (RGC) of the Hong Kong Joint Research Scheme (Project No. N_HKU750/23), HKU seed fund, and the Shenzhen-Hong Kong-Macau Technology Research Programme (Category C project, no. SGDX20230821091501008).

Y.L. acknowledges support from the Research Grants Council of Hong Kong under the General Research Fund (Grant no. 17210520), and the National Natural Science Foundation of China (Grant no. 12272332, 12572194).

**Data Availability Statement**

All data are available in the main text or the supplementary materials.

**References**

[1] B. Geiger, J. P. Spatz, A. D. Bershadsky, Environmental sensing through focal adhesions, *Nature Reviews Molecular Cell Biology* **2009**, 10, 21.10.1038/nrm2593

[2] J. D. Humphrey, E. R. Dufresne, M. A. Schwartz, Mechanotransduction and extracellular matrix homeostasis, *Nature Reviews Molecular Cell Biology* **2014**, 15, 802.10.1038/nrm3896

[3] K. H. Vining, D. J. Mooney, Mechanical forces direct stem cell behaviour in development and regeneration, *Nature Reviews Molecular Cell Biology* **2017**, 18, 728.10.1038/nrm.2017.108

[4] O. Chaudhuri, J. Cooper-White, P. A. Janmey, D. J. Mooney, V. B. Shenoy, Effects of extracellular matrix viscoelasticity on cellular behaviour, *Nature* **2020**, 584, 535.10.1038/s41586-020-2612-2

[5] W. Xie, X. Wei, H. Kang, H. Jiang, Z. Chu, Y. Lin, Y. Hou, Q. Wei, Static and Dynamic: Evolving Biomaterial Mechanical Properties to Control Cellular

Mechanotransduction, *Advanced Science* **2023**, 10, 2204594.https://doi.org/10.1002/advs.202204594

[6] W. Xie, L. Ma, P. Wang, X. Liu, D. Wu, Y. Lin, Z. Chu, Y. Hou, Q. Wei, Dynamic Regulation of Cell Mechanotransduction through Sequentially Controlled Mobile Surfaces, *Nano Letters* **2024**, 24, 7953.10.1021/acs.nanolett.4c01371

[7] F. Martino, A. R. Perestrelo, V. Vinarský, S. Pagliari, G. Forte, Cellular Mechanotransduction: From Tension to Function, *Frontiers in Physiology* **2018**, Volume 9 - 2018.10.3389/fphys.2018.00824

[8] M. Bergert, T. Lendenmann, M. Zündel, A. E. Ehret, D. Panozzo, P. Richner, D. K. Kim, S. J. P. Kress, D. J. Norris, O. Sorkine-Hornung, E. Mazza, D. Poulikakos, A. Ferrari, Confocal reference free traction force microscopy, *Nature Communications* **2016**, 7, 12814.10.1038/ncomms12814

[9] S. S. Hur, J. H. Jeong, M. J. Ban, J. H. Park, J. K. Yoon, Y. Hwang, Traction force microscopy for understanding cellular mechanotransduction, *BMB Rep* **2020**, 53, 74.10.5483/BMBRep.2020.53.2.308

[10] W. J. Polacheck, C. S. Chen, Measuring cell-generated forces: a guide to the available tools, *Nature Methods* **2016**, 13, 415.10.1038/nmeth.3834

[11] M. Dembo, Y.-L. Wang, Stresses at the Cell-to-Substrate Interface during Locomotion of Fibroblasts, *Biophysical Journal* **1999**, 76, 2307.10.1016/S0006-3495(99)77386-8

[12] S. G. Knoll, M. Y. Ali, M. T. Saif, A novel method for localizing reporter fluorescent beads near the cell culture surface for traction force microscopy, *J Vis Exp* **2014**, DOI: 10.3791/5187351873.10.3791/51873

[13] Y. Hou, X. Hu, C. Qian, W. Xie, L. Ma, L. Zhang, X. Han, Y. Tan, Y. Lin, C. Fang, Z. Chu, Topology Outweighs Stiffness: Self-Reinforced Cell Mechanotransduction via Multiaxial Curvature Engineering of Ultrasoft Hydrogels, *ACS Nano* **2026**, DOI: 10.1021/acsnano.5c19367.10.1021/acsnano.5c19367

[14] J. L. Tan, J. Tien, D. M. Pirone, D. S. Gray, K. Bhadriraju, C. S. Chen, Cells lying on a bed of microneedles: An approach to isolate mechanical force, *Proceedings of the National Academy of Sciences* **2003**, 100, 1484.doi:10.1073/pnas.0235407100

[15] L. Schermelleh, A. Ferrand, T. Huser, C. Eggeling, M. Sauer, O. Biehlmaier, G. P. C. Drummen, Super-resolution microscopy demystified, *Nature Cell Biology* **2019**, 21, 72.10.1038/s41556-018-0251-8

[16] L. Wang, Y. Hou, T. Zhang, X. Wei, Y. Zhou, D. Lei, Q. Wei, Y. Lin, Z. Chu, All-Optical Modulation of Single Defects in Nanodiamonds: Revealing Rotational and Translational Motions in Cell Traction Force Fields, *Nano Letters* **2022**, 22, 7714.10.1021/acs.nanolett.2c02232

[17] J. M. Gere, S. Timoshenko, Mechanics of Materials. ed, *Boston, MA: PWS* **1997**,

[18] T. Zhang, G. Pramanik, K. Zhang, M. Gulka, L. Wang, J. Jing, F. Xu, Z. Li, Q. Wei, P. Cigler, Z. Chu, Toward Quantitative Bio-sensing with Nitrogen-Vacancy Center in Diamond, *ACS Sens* **2021**, 6, 2077.10.1021/acssensors.1c00415

[19] R. Schirhagl, K. Chang, M. Loretz, C. L. Degen, Nitrogen-vacancy centers in diamond: nanoscale sensors for physics and biology, *Annu Rev Phys Chem* **2014**, 65, 83.10.1146/annurev-physchem-040513-103659

[20] M. W. Doherty, N. B. Manson, P. Delaney, F. Jelezko, J. Wrachtrup, L. C. L. Hollenberg, The nitrogen-vacancy colour centre in diamond, *Physics Reports* **2013**, 528, 1.https://doi.org/10.1016/j.physrep.2013.02.001

[21] J. M. Taylor, P. Cappellaro, L. Childress, L. Jiang, D. Budker, P. R. Hemmer, A. Yacoby, R. Walsworth, M. D. Lukin, High-sensitivity diamond magnetometer with nanoscale resolution, *Nature Physics* **2008**, 4, 810.10.1038/nphys1075

[22] P. Maletinsky, S. Hong, M. S. Grinolds, B. Hausmann, M. D. Lukin, R. L. Walsworth, M. Loncar, A. Yacoby, A robust scanning diamond sensor for nanoscale imaging with

single nitrogen-vacancy centres, *Nat Nanotechnol* **2012**, 7, 320.10.1038/nnano.2012.50

[23] S. Hong, M. S. Grinolds, L. M. Pham, D. Le Sage, L. Luan, R. L. Walsworth, A. Yacoby, Nanoscale magnetometry with NV centers in diamond, *MRS Bulletin* **2013**, 38, 155.10.1557/mrs.2013.23

[24] K. Xia, C. F. Liu, W. H. Leong, M. H. Kwok, Z. Y. Yang, X. Feng, R. B. Liu, Q. Li, Nanometer-precision non-local deformation reconstruction using nanodiamond sensing, *Nat Commun* **2019**, 10, 3259.10.1038/s41467-019-11252-3

[25] T. P. M. Alegre, C. Santori, G. Medeiros-Ribeiro, R. G. Beausoleil, Polarization-selective excitation of nitrogen vacancy centers in diamond, *Physical Review B—Condensed Matter and Materials Physics* **2007**, 76, 165205

[26] M. Li, H. Yuan, P. Fan, S. Wang, J. Shen, L. Xu, Simultaneous vector magnetometry based on fluorescence polarization of NV centers ensemble in diamond, *Applied Physics Letters* **2024**, 125.10.1063/5.0220694

[27] A. Öchsner, *Classical beam theories of structural mechanics*, Springer, **2021**.

[28] R. W. Style, R. Boltyanskiy, G. K. German, C. Hyland, C. W. MacMinn, A. F. Mertz, L. A. Wilen, Y. Xu, E. R. Dufresne, Traction force microscopy in physics and biology, *Soft Matter* **2014**, 10, 4047.10.1039/C4SM00264D

[29] P. Roca-Cusachs, V. Conte, X. Trepat, Quantifying forces in cell biology, *Nature Cell Biology* **2017**, 19, 742.10.1038/ncb3564

[30] L. Shanahan, S. Belser, J. W. Hart, Q. Gu, J. R. E. Roth, A. Mechnich, M. Högen, S. Pal, T. Mitchell, D. Jordan, E. A. Miska, M. Atatüre, H. S. Knowles, Q-BiC: A Biocompatible Integrated Chip for in vitro and in vivo Spin-Based Quantum Sensing, *PRX Life* **2025**, 3, 013016.10.1103/PRXLife.3.013016

The table of contents entry should be 50–60 words long and should be written in the present tense. The text should be different from the abstract text.

Cellular forces are traditionally inferred from displacement measurements that suffer from diffraction and modeling limitations. This work replaces displacement with rotational angle for the fundamental mechanical readout. Using diamond quantum sensing, three-dimensional pillar rotations are directly measured, enabling accurate and multidimensional cellular force quantification and redefining traction force microscopy toward angle-resolved biomechanical sensing.

**From Displacement to Angle: Diamond-Based 3D Rotation Sensing for High-Precision Cellular Force Measurement**

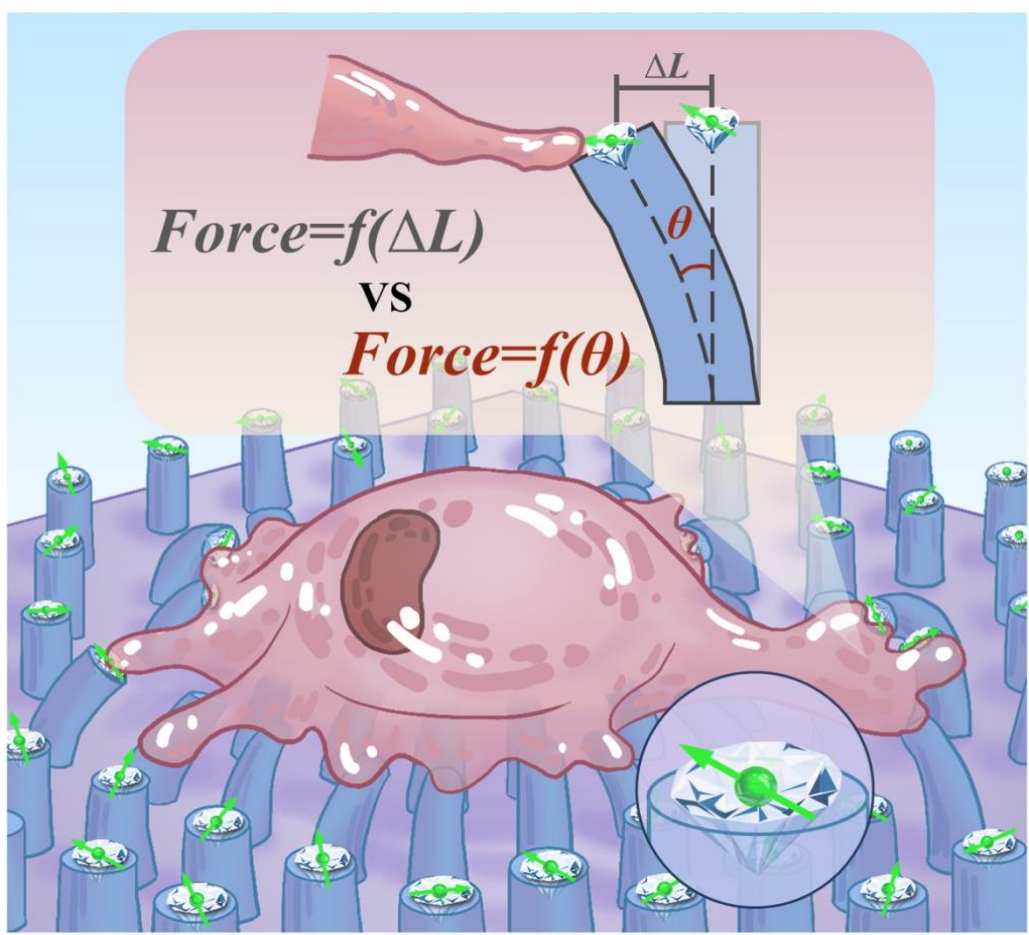

Supporting Information

**From Displacement to Angle: Diamond-Based 3D Rotation Sensing for High-Precision Cellular Force Measurement**

*Linjie Ma[1#], Bicong Wang[2#], Tai Nam Yip[1], Yong Hou[1]\*, Yuan Lin[2]\*, and Zhiqin Chu[1,3]\**

1. Department of Electrical and Electronic Engineering, The University of Hong Kong, Hong Kong SAR
2. Department of Mechanical Engineering, The University of Hong Kong, Hong Kong SAR
3. School of Biomedical Sciences, The University of Hong Kong, Hong Kong SAR

\*Corresponding authors:
Dr. Yong Hou, Email: houyong@eee.hku.hk
Prof. Dr. Yuan Lin, Email: ylin@hku.hk
Prof. Dr. Zhiqin Chu, E-mail: zqchu@eee.hku.hk

[#] These authors made equal contributions to this work.

**Supporting Information Note 1: Model of bending pillar**

**Deformation of the cantilever beam**

When a cantilever beam is applied a concentrated force *P* at the free end, as described by **Figure S1**, its curvature is given by the equation:

$$\frac{1}{\rho} = \frac{d\theta}{ds} \tag{S1}$$

where $\rho$ is the radius of curvature of the neutral surface. Additionally, if the beam in pure bending is linearly elastic and follows Hooke's law, the curvature is:

$$\frac{1}{\rho} = \frac{M(s)}{EI} = \frac{P[x(L)-x(s)]}{EI}. \tag{S4}$$

The sign conventions to be used with **Equation S2** are as follows: (1) the *x*-axis is positive to the right; (2) the rotational angle $\theta$ is positive when clockwise from *x*-axis; (3) the deflection *w* is positive downward; (4) the bending moment is positive when it produces stretching in the upper part of the beam.

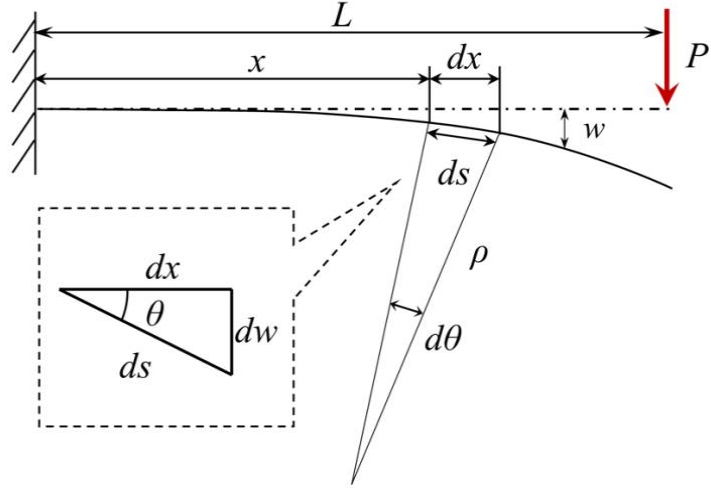

**Figure S1.** Bending of a beam.

The classical Euler-Bernoulli beam theory assumed that the deformations are very small and the plane sections remain plane. Combing **Equation S1** and **Equation S2**, we can get:

$$\frac{d\theta}{ds} = \frac{d\theta}{dx}\frac{dx}{ds} = \frac{d\theta}{dx}\frac{1}{\sqrt{1+\left(\frac{dw}{dx}\right)^2}} = \frac{P(L-x)}{EI}. \qquad (S3)$$

From the inset of Figure S1, we see that:

$$\theta = \arctan\frac{dw}{dx}, \quad \frac{d\theta}{dx} = \frac{d^2w}{dx^2}\frac{1}{1+\left(\frac{dw}{dx}\right)^2}, \qquad (S5)$$

With the assumption of small deflections, $1+(dw/dx)^2 \approx 1$, we obtain [1]:

$$\frac{d^2w_{EB}}{dx^2} \approx \frac{d\theta_{EB}}{dx} \approx \frac{P(L-x)}{EI}. \qquad (S6)$$

Thus, the rotational angle $\theta_{EB}(L) = PL^2/(2EI)$ and deflection $w_{EB}(L) = PL^3/(3EI)$.

Here, to estimate the errors brought by $1+(dw/dx)^2 \approx 1$, we rewrite **Equation S3** by substituting $dw/dx = \tan\theta$:

$$|\cos\theta|d\theta = \frac{P}{EI}(L-x)dx. \qquad (S7)$$

Typically, the maximum angle of rotation at $x = L$ is smaller than $\pi/2$, resulting $|\cos\theta(x)| = \cos\theta(x)$. Considering the boundary condition $\theta(0) = 0$, we can obtain:

$$\theta_{IF} = \arcsin\left[\frac{P}{EI}\left(Lx - \frac{x^2}{2}\right)\right] = \sum_{n=0}^{\infty}\frac{(2n)!}{4^n(n!)^2(2n+1)}\frac{P^{2n+1}}{E^{2n+1}I^{2n+1}}\left(Lx - \frac{x^2}{2}\right)^{2n+1}. \qquad (S8)$$

It is noted that the expression of **Equation S7** reduces to the classical Euler–Bernoulli solutions when the terms with $n > 0$ are ignored. The deflection curve:

$$w_{IF} = \int_0^x \tan\theta_{IF}\, dx = \int_0^x \frac{\sin\theta_{IF}}{\sqrt{1-\sin^2\theta_{IF}}}dx.$$

The integral of the above equation yields an incomplete elliptic integral, which cannot be expressed in terms of elementary functions. Thus, we use the series expansion for the tangent function:

$$w_{IF} = \int_0^x \tan\theta_{IF}\,dx = \int_0^x \left(\theta_{IF} + \frac{1}{3}\theta_{IF}^3 + \cdots\right)dx$$

$$= \int_0^x \left[\left(\frac{P}{EI}\left(Lx - \frac{x^2}{2}\right) + \frac{1}{6}\frac{P^3}{E^3I^3}\left(Lx - \frac{x^2}{2}\right)^3 + \cdots\right) + \frac{1}{3}\left(\frac{P}{EI}\left(Lx - \frac{x^2}{2}\right) + \cdots\right)^3 + \cdots\right]dx$$

$$= \frac{P}{EI}\left(\frac{Lx^2}{2} - \frac{x^3}{6}\right) + \frac{P^3}{2E^3I^3}\left(\frac{1}{4}L^3x^4 - \frac{3}{10}L^2x^5 + \frac{3}{24}Lx^6 - \frac{x^7}{56}\right) + \cdots. \quad (S9)$$

Through the above derivation, it can be shown that approximating $1+(dw/dx)^2$ by 1 gives rise to errors in rotational angle and deflection at $x = L$ of:

$$e_\theta = \left|\frac{\theta_{EB} - \theta_{IF}}{\theta_{IF}}\right| \approx \frac{1}{1 + \frac{(24E^2I^2)}{(P^2L^4)}}$$

$$e_w = \left|\frac{w_{EB} - w_{IF}}{w_{IF}}\right| \approx \frac{1}{1 + \frac{(35E^2I^2)}{(3P^2L^4)}} \quad (S10)$$

It suggests that, within the framework of Euler-Bernoulli beam theory (Equation S3), the deviation in the linear force-displacement relationship is more pronounced than that in the force-angle relationship since $e_\theta < e_w$ even when the applied force is small (**Figure 1c**)

However, even **Equation S7** and **S8** compensate for the error arising from the approximation $1+(dw/dx)^2 \approx 1$, they remain relatively high accuracy only under small deformations. This is due to a more subtle approximation, $x(s) \approx s$, introduced during the derivation of Equation S3. Clearly, this approximation loses its reliability under large deformations. We can return to Equation S1 and S2 and re-derive the relationship between load and deformation. By taking the derivative with respect to $s$, we obtain a second-order ordinary differential equation as follow:

$$\frac{d^2\theta}{ds^2} = -\frac{P}{EI}\frac{dx}{ds} = -\frac{P}{EI}\cos\theta, \quad (S11)$$

which can be solved as:

$$\frac{1}{2}\left(\frac{d\theta}{ds}\right)^2 = \frac{P}{EI}(\sin\theta_L - \sin\theta)$$

with $\theta_L$ being the realistic rotational angle of the free end of the beam. The above expression implies the implicit relationship between $\theta$ and $P$:

$$\int_0^{\theta_L} \frac{d\theta}{\sqrt{\sin\theta_L - \sin\theta}} = \int_0^L \sqrt{\frac{2P}{EI}}\,ds. \quad (S12)$$

By introducing substitution $\sin\theta = 2k^2\sin^2\psi - 1$ and $2k^2 = \sin\theta_L + 1$, Eq. S11 can be derived into:

$$\int_{\arcsin\frac{1}{\sqrt{2}k}}^{\frac{\pi}{2}} \frac{1}{\sqrt{2}k\sqrt{1-\sin^2\psi}} \frac{2k\cos\psi\,d\psi}{\sqrt{1-k^2\sin^2\psi}} = \int_{\arcsin\frac{1}{\sqrt{2}k}}^{\frac{\pi}{2}} \frac{\sqrt{2}d\psi}{\sqrt{1-k^2\sin^2\psi}} = \sqrt{\frac{2P}{EI}}L.$$

which is usually expressed in terms of elliptic integrals of the first kind [2]:

$$F(k) - F(k, \psi_0) = \sqrt{\frac{P}{EI}} L, \psi_0 = \arcsin\frac{1}{\sqrt{2}k}, k = \sqrt{\frac{\sin\theta_L + 1}{2}}. \tag{S13}$$

where $F(k)$ and $F(k, \psi_0)$ respectively represent the complete and incomplete elliptic integral of the first kind.

**Tilting of the pillar base**

The preceding section derived the deformation of a beam with a fixed end at $x = 0$. However, the pillar arrays made of polydimethylsiloxane (PDMS) are monolithic with the substrate. When a lateral force is applied to the top of the pillar, the pillar base, where they connect to the substrate, deforms under the action of the torque $PL$ (**Figure S2a**) [3-4]. The tilting angle of the pillar base can be directly calculated by $\arctan(u_A/R)$. The stress distribution at the bottom cross-section is given as follows:

$$\sigma(y, z) = -\frac{PLy}{I}, \quad y^2 + z^2 \leq R^2. \tag{S14}$$

For a soft elastic half-space, the vertical displacement caused by a concentrated force q is given by the Boussinesq solution [5]:

$$u = \frac{1-v^2}{\pi E}\frac{q}{r}, \tag{S15}$$

where $r$ is the distance from the concentrated force, $v$ is the Poisson ratio. Thus, the vertical displacement at $B$ ($y_B$, $z_B$) (**Figure S2b**, $y_B \leq 0$) due to $\sigma(y, z)$ with area $dydz$:

$$u(y,z) = -\frac{1-v^2}{\pi E}\frac{PLy}{I}\frac{dydz}{\sqrt{(y-y_B)^2 + (z-z_B)^2}}. \tag{S16}$$

Here, we only consider $z_B = 0$ for simplicity. Introducing substitution $y = y_B + r\cos\varphi$, $z = r\sin\varphi$:

$$u(r,\varphi) = -\frac{1-v^2}{\pi E}\frac{PL}{I}(y_B + r\cos\varphi)drd\varphi. \tag{S17}$$

Then the total displacement at $A$ is:

$$u_B = -\frac{1-v^2}{\pi E}\frac{PL}{I}\int_0^\pi \int_{r_1}^{r_2}(y_B + r\cos\varphi)dr\,d\varphi, \tag{S18}$$

where $r_{1,2} = (-y_B\cos\varphi) \mp \sqrt{R^2 - y_B^2\sin^2\varphi}$, thus,

$$u_B = -\frac{1-v^2}{\pi E}\frac{PL}{I}\int_0^\pi -2y_B\sin^2\varphi\sqrt{R^2 - y_B^2\sin^2\varphi}\,d\varphi.$$

Introducing substitution $h = -y_B\sin\varphi$,

$$u_B = -\frac{1-v^2}{\pi E}\frac{PL}{I}\cdot 2\int_0^{-y_B}\frac{2y_B\sqrt{R^2-h^2}}{\sqrt{y_B^2-h^2}}dh = -\frac{1-v^2}{\pi E}\frac{PL}{I}\frac{4}{y_B}\int_0^{-y_B}\frac{h^2\sqrt{R^2-h^2}}{\sqrt{y_B^2-h^2}}dh$$

$$= -\frac{1-v^2}{\pi E}\frac{PL}{I}\frac{4}{y_B}\left[\int_0^{-y_B}\frac{R^2h^2}{\sqrt{R^2-h^2}\sqrt{y_B^2-h^2}}dh - \int_0^{-y_B}\frac{h^4}{\sqrt{R^2-h^2}\sqrt{y_B^2-h^2}}dh\right]$$

$$= -\frac{1-v^2}{\pi E}\frac{PL}{I}\frac{4}{3}\left[\left(\frac{R^3}{y_B}-Ry_B\right)F\left(\frac{|y_B|}{R}\right)+\left(2Ry_B-\frac{R^3}{y_B}\right)E\left(\frac{|y_B|}{R}\right)\right],$$

where $F(|y_B|/R)$ and $E(|y_B|/R)$ are respectively the first kind and second kind complete elliptic integrals with modulus ($|y_B|/R$). The vertical displacement at the edge point $A$ is:

$$u_A = \frac{4(1-v^2)PLR^2}{3\pi EI}E(1) = \frac{4(1-v^2)PLR^2}{3\pi EI}. \tag{S19}$$

Hence, the tilting angle of the pillar base $\theta_T$ can be calculated by:

$$\theta_T = arctan\frac{u_A}{R} \approx \frac{4(1-v^2)PLR}{3\pi EI}. \tag{S20}$$

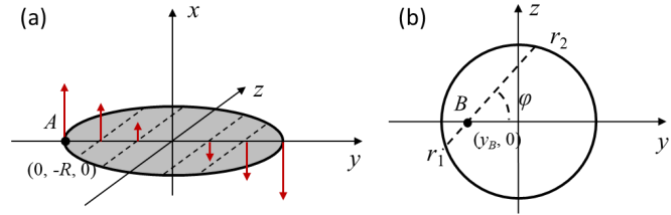

**Figure S2.** a) Stress distribution at the pillar bottom. b) Displacement at an internal point $B$.

In practice, the rotation angle of the pillar top measured by nanodiamonds $\theta_M$ comprises contributions from bending of the beam $\theta_L$ and tilting of the base $\theta_T$. We can estimate the proportion of $\theta_L$ relative to $\theta_M$ through the Euler–Bernoulli beam theory:

$$C_\theta = \frac{\theta_L}{\theta_M} = \frac{\frac{PL^2}{2EI}}{\frac{PL^2}{2EI}+\frac{4(1-v^2)PLR}{3\pi EI}} = \frac{1}{\frac{4(1-v^2)D}{3\pi}\frac{1}{L}+1}. \tag{S21}$$

That is, the measured angle $\theta_M$ will be first multiplied by a correction factor $C_\theta$ before being applied to the **Equation S12** for force calculation.

**Numerical simulation**

We conducted finite element simulations with circular pillar height of 6 μm, Poisson ratio 0.47 and Young's Modulus 1.49 MPa. Diameters were set as 1.5, 2 and 3 μm, respectively. The substrate measured $10^3$ μm$^3$ with fixed boundaries at its bottom and surrounding. The traction force was respectively applied as homogenous facial load $p$ to the top surface of the pillar. The mesh was extra refined and geometric nonlinearity was included. The rotational angle of top surfaces was calculated from the vertical displacements of edge points.

# Supporting Information Note 2: Theoretical Model for Determining NV Orientation via ODMR-LPM Hybrid Method

## NV Orientation and Determination of Rotational Motion

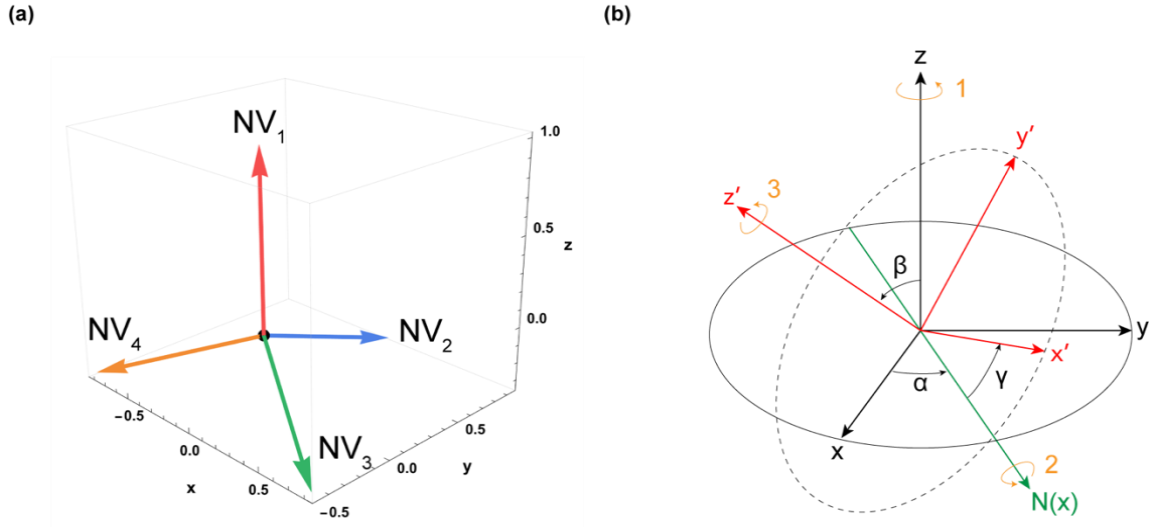

**Figure S3.** a) The four crystallographic orientations of NV centers in the reference state. b) The Euler angle following the ZXZ convention.

In a single-crystalline diamond, the NV center can adopt one of four crystallographic axes, corresponding to: $(111)$, $(\bar{1}\bar{1}1)$, $(\bar{1}1\bar{1})$, and $(1\bar{1}\bar{1})$, expressed in Miller indices. These four directions are symmetrically equivalent in the diamond lattice [6].

To fully determine the three-dimensional rotation of a nanodiamond, we track the relative orientations of the ensemble of NV centers inside. We define a reference diamond orientation $NV^{ref}$ (denoted as the reference frame) in which the $(111)$-oriented NV center is aligned with the laboratory Z-axis, and the $(\bar{1}\bar{1}1)$-oriented NV lies within the Y-Z plane (**Figure S3a**).

In this reference configuration, the unit vectors corresponding to the four NV axes are:

$$NV_1^{ref} = (0, 0, 1),\ NV_2^{ref} = \left(0, \frac{2\sqrt{2}}{3}, -\frac{1}{3}\right),$$

$$NV_3^{ref} = \left(\frac{\sqrt{6}}{3}, -\frac{\sqrt{2}}{3}, -\frac{1}{3}\right),\ NV_4^{ref} = \left(\frac{\sqrt{6}}{3}, -\frac{\sqrt{2}}{3}, -\frac{1}{3}\right).$$

The orientation of the diamond in the laboratory frame is parameterized using Euler angles $(\alpha, \beta, \gamma)$ following the ZXZ convention (shown in **Figure S3b**, all the following Euler angle follows ZXZ convention). This means the rotation is composed of three sequential steps:

1. A rotation by angle $\alpha$ around the laboratory z-axis, yielding an intermediate frame $(x', y', z')$,
2. A rotation by angle $\beta$ around the new $x'$-axis, producing frame $(x'', y'', z'')$,
3. A final rotation by angle $\gamma$ around the new $z''$-axis, resulting in the final orientation of the diamond.

The rotation matrix of the operation is [7]

Rotation about z-axis:

$$R_z(\theta) = \begin{pmatrix} Cos(\theta) & -Sin(\theta) & 0 \\ Sin(\theta) & Cos(\theta) & 0 \\ 0 & 0 & 1 \end{pmatrix}$$

Rotation about x-axis:

$$R_x(\theta) = \begin{pmatrix} 1 & 0 & 0 \\ 0 & Cos(\theta) & -Sin(\theta) \\ 0 & Sin(\theta) & Cos(\theta) \end{pmatrix}$$

The rotation matrix of the Euler angle $(\alpha, \beta, \gamma)$:

$$R(\alpha, \beta, \gamma) = R_z(\alpha)R_x(\beta)R_z(\gamma)$$

The corresponding NV orientation is:

$$NV'_i = R(\alpha, \beta, \gamma)NV_i^{ref}$$

By applying this composite rotation to the reference NV vectors, one can compute the expected orientations of the NV centers under arbitrary diamond rotation. Because all these 4 NV axis are symmetrically equivalent in the diamond lattice, they can be randomly chosen.

Assume that the Euler angles of the diamond in two distinct orientations, denoted as state 1 and state 2, are given by $(\alpha_1, \beta_1, \gamma_1)$ and $(\alpha_2, \beta_2, \gamma_2)$ respectively. Let $R_1 = R(\alpha_2, \beta_2, \gamma_2)$ and $R(\alpha_2, \beta_2, \gamma_2)$ represent the corresponding rotation matrices constructed using the ZXZ Euler convention.

The relative rotation matrix $R_{rel}$, which describes the rotation of the diamond from state 1 to state 2, is given by:

$$R_{rel} = R_2 \cdot R_1^{-1}$$

This matrix transforms vectors expressed in the reference frame of state 1 into their corresponding representations in state 2. It can be directly applied to the NV orientation vectors to compute their new orientations after the diamond rotation, thereby enabling precise determination of the diamond's three-dimensional rotational motion.

Given the relative rotation matrix $R_{rel}$, the corresponding rotation axis and rotation angle can be extracted to provide an intuitive geometric interpretation of the diamond's reorientation. The rotation angle $\theta$ is calculated using the trace of the rotation matrix [7]:

$$\theta = Cos^{-1}\left(\frac{Tr(R_{rel}) - 1}{2}\right)$$

Where $Tr(R_{rel})$ denotes the trace (the sum of the diagonal elements) of the rotation matrix. The unit rotation axis $u = (u_x, u_y, u_z)$ can be obtained from the off-diagonal elements of $R_{rel}$ as follows:

$$u_x = \frac{R_{32} - R_{23}}{2Sin\theta}$$

$$u_y = \frac{R_{13} - R_{31}}{2Sin\theta}$$

$$u_z = \frac{R_{21} - R_{12}}{2Sin\theta}$$

where $R_{ij}$ denotes the element in the $i$-th row and $j$-th column of $R_{rel}$. These expressions are valid when $\theta \neq 0, \pi$. In degenerate cases (e.g., zero or 180° rotation), axis extraction requires special handling; however, such cases are very rare in practice, so we omit the related discussion.

**Modeling of ODMR-LPM Hybrid Method for NV Axis Identification**

Here, we present the theoretical basis of using the ODMR-LPM hybrid method to determine the orientation of the NV axis and the corresponding Euler angles of the nanodiamond.

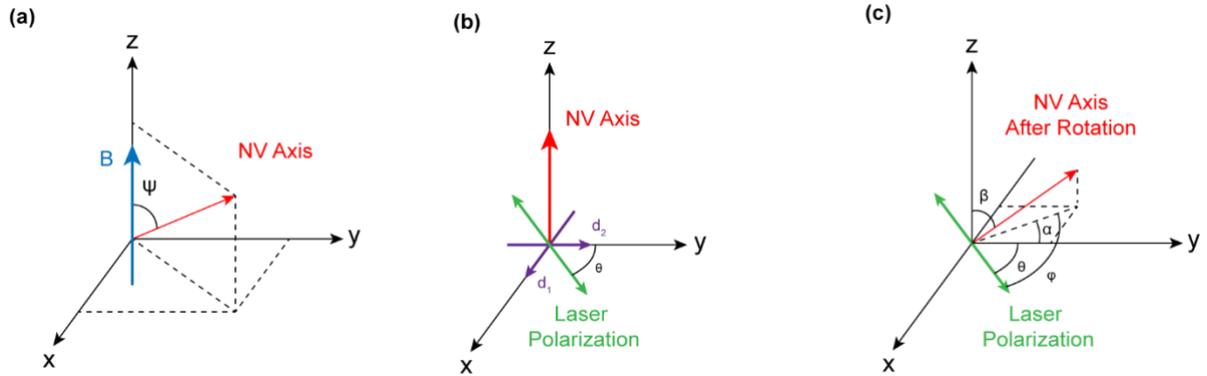

**Figure S4.** a) The angle between the NV axis and the magnetic field. b) Schematic of the NV axis and the corresponding dipoles. The laser polarization direction is also indicated in the figure. c) Schematic of the angle between the projection of the NV axis on the x-y plane and laser politization direction after rotation.

The angle between the external magnetic field can be measured from the frequency shift of the ODMR spectrum [8]. The ground state of the NV center is a spin triplet state. The Hamiltonian of it under an external magnetic field $\vec{B}$ can be written as [9]:

$$H = D(\vec{NV} \cdot \hat{S})^2 + \gamma_e \vec{B} \cdot \hat{S}$$

Where $D \approx 2.87\ GHz$ is the zero-field splitting between the $m_s = 0$ state and $m_s = \pm 1$ states, $\hat{S}$ is the spin operator, and $\gamma = 28\ MHz/mT$ is the electron gyromagnetic ratio. The resonant frequency can be written as [8]:

$$f^\pm = D + \frac{3\gamma_e^2 B^2}{2D} sin^2\psi \pm \gamma_e B cos\psi \sqrt{1 + \frac{\gamma_e^2 B^2}{4D^2} tan^2\psi\, sin^2\psi}$$

Where $\psi = arccos\frac{\vec{NV} \cdot \vec{B}}{B}$ is the angle between the NV axis and the extremal magnetic field. When the orientation and magnitude of the extremal magnetic field B is given, the angle $\psi$ can be fully determined by fitting the resonant frequency. In fact, in our experiment, because the magnetic field direction is vertical (the validation is shown in the SI), the calculated angle $\psi$ is the angle between the NV center and the Z-axis (as shown in **Figure S4a**). However, this formula also shows that the rotation around the z-axis can not be solved because it will not induce any change of the $\psi$. This can be solved by applying another method that can measure the horizontal component.

To resolve the ambiguity in NV axis identification, we use an excitation polarization–dependent fluorescence method. It is well-established that the excitation efficiency of an NV center is governed by the coupling strength between the electric field vector of the excitation laser and the dipole moments associated with the NV's electron orbitals[10].

Each NV center possesses two orthogonal dipole vectors, $\vec{d_1}$ and $\vec{d_2}$, both lying in the plane perpendicular to the NV axis[11-13]. The excitation probability for each dipole is proportional to the square of the projection of the laser's electric field $\vec{\varepsilon}$ onto the dipole vector:

$$\Gamma \propto |\vec{\varepsilon} \cdot \vec{d}|^2$$

The linearly polarized laser has a polarization angle $\theta$ measured from the Y-axis (as shown in **Figure S4b**), and the corresponding electric field vector is expressed as:

$$\vec{\varepsilon} = E \begin{pmatrix} sin\theta \\ cos\theta \\ 0 \end{pmatrix}$$

Assuming an NV center aligned along the [001] crystallographic direction, its two dipoles in the initial configuration (before rotation) can be represented as [10, 12]:

$$\vec{d_1} = d \begin{pmatrix} 1 \\ 0 \\ 0 \end{pmatrix}, \vec{d_2} = d \begin{pmatrix} 0 \\ 1 \\ 0 \end{pmatrix}$$

After applying a general rotation parameterized by the Euler angles $(\alpha, \beta, \gamma)$ following the ZXZ convention, the rotated dipole vectors become:

$$\vec{d_1'} = R_z(\alpha) R_x(\beta) R_z(\gamma) \cdot \vec{d_1}$$
$$\vec{d_2'} = R_z(\alpha) R_x(\beta) R_z(\gamma) \cdot \vec{d_2}$$

Since the fluorescence intensity $I$ is proportional to the total excitation probability, it follows that:

$$I \propto \Gamma_{d_1'} + \Gamma_{d_2'}$$
$$\propto |\vec{\varepsilon} \cdot \vec{d_1'}|^2 + |\vec{\varepsilon} \cdot \vec{d_2'}|^2$$

Under further simplification, the intensity can be expressed as:

$$I \propto dE(1 - \sin^2\beta \cos^2\varphi)$$

$$I \propto 1 - \sin^2\beta \cos^2\varphi$$

where $\varphi = \alpha + \theta$. Here, as shown in **Figure S4c**, $\alpha$ represents the azimuthal angle of the NV axis in the laboratory frame (rotation around the Z-axis, counterclockwise), and $\theta$ is the polarization angle of the incident laser with respect to the Y-axis (clockwise). Therefore, $\varphi$ quantifies the angle between the laser's polarization direction and the projection of the NV axis onto the transverse (xy) plane. Meanwhile, $\beta$ denotes the polar angle between the NV axis and the laser propagation direction (assumed to be the Z-axis).

This relationship reveals that the fluorescence intensity reaches a maximum when $\varphi = 0$, which corresponds to the situation that when the laser polarization is aligned with the projected NV axis. Thus, by measuring the intensity variation as a function of the laser polarization angle, the orientation of a single NV axis can be precisely determined.

However, in nanodiamonds that contain multiple NV centers with different crystallographic orientations, the fluorescence signals from different NVs are superimposed. To resolve this, an external magnetic field can be applied to lift the degeneracy of the NV spin states [14]. As a result, the ODMR frequencies of NV centers with different orientations are split due to the magnetic field's different projections along their respective axes.

Because microwave excitation induces spin transitions at specific resonant frequencies, the ODMR contrast (the difference in fluorescence intensity between microwave-on and microwave-off states) will be selectively modulated only by the NV center that is on resonance. Consequently, even in a multi-NV system, the angular dependence of the ODMR contrast at a given frequency reflects the orientation of a specific NV axis. In this way, the projection direction of an individual NV can still be determined.

**Fitting Method**

As established in the previous section, the orientation of any ND can be regarded as a rotation from a reference configuration orientation $NV^{ref}$, characterized by a set of Euler $(\alpha, \beta, \gamma)$. Specifically, for $NV_1^{ref}$, we have:

$$NV_0' = R(\alpha, \beta, \gamma) NV_0^{ref} = \begin{pmatrix} sin(\alpha)sin(\beta) \\ -cos(\alpha)sin(\beta) \\ cos(\beta) \end{pmatrix}$$

Based on this formula, we can tell its projection on the x-y plane forms an angle of $(\frac{\pi}{2} - \alpha)$, which is only related to $\alpha$, and the change in $\beta$ and $\gamma$ will not influence it. Based on this, by measuring its projection on x-y plane, the value of alpha can be fully determined. That can be measured using the LPM method.

The angle $\beta$ and $\gamma$ can be measured based on the ODMR method. In fact, because the Euler angle $\alpha$ represents the rotation around the Z-axis, the change of $\alpha$ will not cause any change in the ODMR spectrum. Based on this, we can split the horizontal and vertical rotation.

To determine the angle $\beta$ and $\gamma$, we directly fit the ODMR spectrum of the ND with the following function [9]:

$$S(f) = c - \sum_i a_i \left[ \frac{\Delta f^2}{4(f - f_i^-)^2 + \Delta f^2} + \frac{\Delta f^2}{4(f - f_i^+)^2 + \Delta f^2} \right]$$

Where $c$ is the baseline, $a_i$ is the contrast of NV centers along the i-th direction, $\Delta f$ is the FWHM. The $f_i^\pm$ is determined from the formula [8]:

$$f_i^\pm = D + \Delta D + \frac{3\gamma_e^2 B^2}{2(D + \Delta D)} \sin^2\psi_i \pm \gamma_e B \cos\psi_i \sqrt{1 + \frac{\gamma_e^2 B^2}{4(D + \Delta D)^2} \tan^2\psi_i \sin^2\psi_i}$$

Where $\Delta D$ is the shift of the zero-field splitting, and:

$$\psi_i = \arccos \frac{\overrightarrow{NV_i} \cdot \vec{B}}{B}$$

$$\overrightarrow{NV_i} = R(\alpha, \beta, \gamma) \overrightarrow{NV_i^{ref}}$$

We use the least-square fitting method, and the fitting parameters are Euler angle $\beta$ and $\gamma$, the baseline $c$, the contrasts $a_i$, shift of the zero-field splitting $\Delta D$, and the FWHM $\Delta f$.

In this way, the Euler angle $\alpha, \beta$ and $\gamma$ can be fully determined. Then the rotation of the ND can be solved based on measurement result of the states.

**Supporting Information Note 3: Measurement Setup**

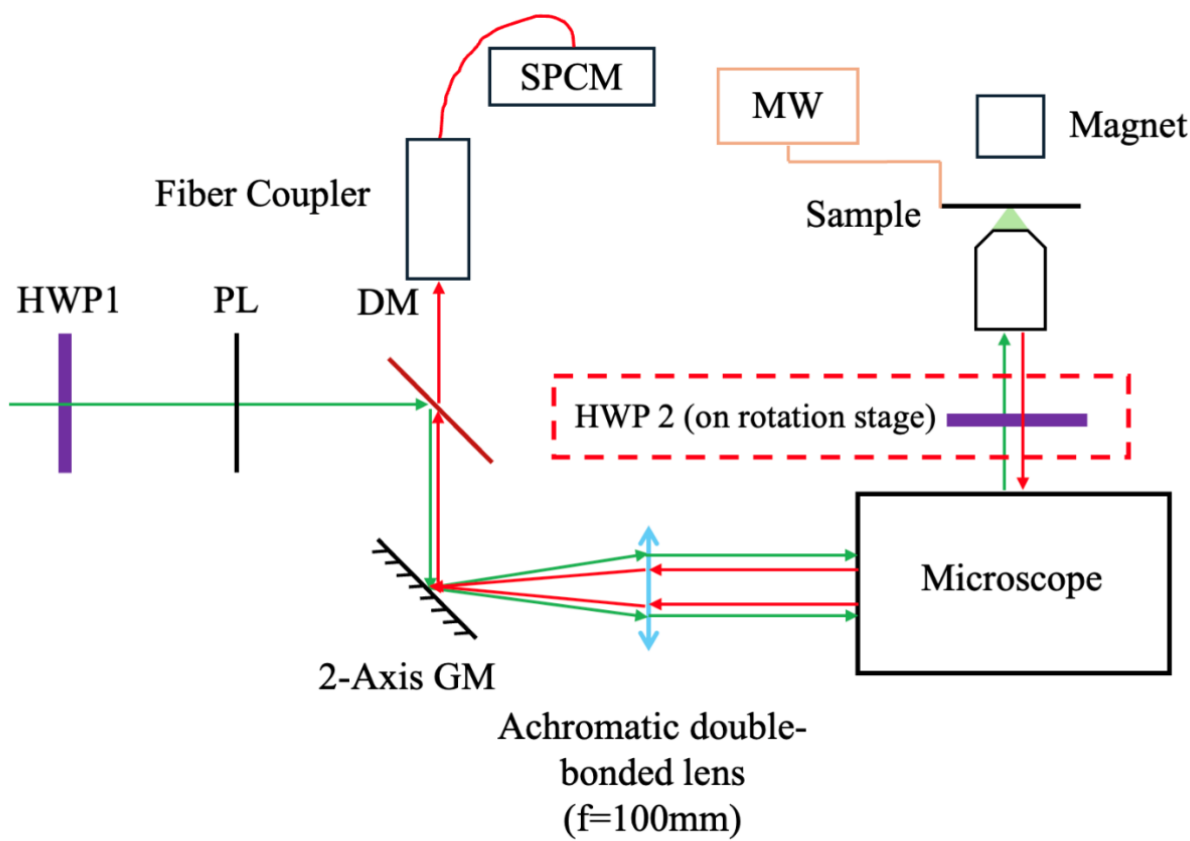

**Figure S5.** Schematic of the measurement setup. HWP: half-wave plate, PL: Polarizer, DM: dichroic mirror, GM: galvo mirror, SPCM: single photon counting module, MW: microwave.

The home-built confocal microscope system used in this study is based on a Nikon Ti2 microscope body. A 532 nm linearly polarized green laser serves as the excitation source. The laser first passes through a half-wave plate (HWP1; HWP25-532A-M, LBTEK) to adjust its polarization orientation. This adjustment is necessary because reflective optical elements can disturb the polarization state unless the incident angle is normal incident light or pure s- or p-polarized light [15]. After that, the beam passes through a polarizer (PL; FLP25-VIS-M, LBTEK) to enhance the degree of polarization. A 560 nm long-pass dichroic mirror (DM; FF560-Di01-25x36, Semrock) is then used to separate the 532 nm excitation light from the NV center fluorescence, which primarily lies in the red spectral range (>637 nm). The laser beam is then directed onto a two-axis galvanometric mirror (GVS012, Thorlabs) for beam scanning. The galvo mirror is controlled via analog signals generated by a data acquisition card (PCIe-6321, National Instruments). An achromatic doublet lens (f = 100 mm, MAD508-A, LBTEK) is positioned before the side port of the microscope and acts as the first lens in a 4f optical system.

Its front focal plane is aligned with the center of two galvanometric mirrors, and its back focal plane coincides with the intermediate image plane of the microscope (typically where a camera sensor is placed). The second lens in the 4f system is the internal tube lens of the Nikon microscope, with a focal length of 200 mm. To manipulate the polarization of the laser, a second half-wave plate (HWP2; HWP25-532A-M, LBTEK) is mounted on a motorized rotation stage (EM-RP60, LBTEK) and placed between the tube lens and the objective. The excitation beam is then focused onto the sample via an air objective with a numerical aperture of 0.95 (UPLANXAPO40X, Olympus).

Fluorescence emitted by the sample is coupled into an optical fiber using a fiber coupler (MBT613D, Thorlabs) with a 10× objective (PLN10X/0.25, Olympus). The signal is detected by a single-photon counting module (SPCM-AQRH-44-FC, EXCELITAS), which is an avalanche photodiode (APD).

For spin manipulation, a microwave signal is generated by a microwave source (SynthNV Pro, Windfreak Technologies), amplified by a high-power amplifier (ZHL-16W-43-S+, Mini-Circuits), and delivered to the sample via an omega-shaped coplanar gold waveguide patterned on the cover glass.

An external magnetic field is applied to the sample using a permanent magnet placed above the objective lens.

**Supporting Information Note 4: Validation of the direction of the magnetic field**

To verify that the magnetic field applied by the permanent magnet was oriented along the vertical axis, we performed ODMR measurements on a bulk diamond with a [100]-cut. In such a crystal, the NV centers are theoretically oriented along four different body diagonals. If the external magnetic field is aligned vertically, this geometry leads to only two resonance peaks in the ODMR spectrum. Using this criterion, we try to align the direction of the magnetic field. After the alignment of the magnetic field, to further validate its direction, we rotated the bulk diamond horizontally and measured the ODMR spectra, fitting the measured ODMR spectrum with the method that we proposed before. The fitting result gives the $\beta$ and $\gamma$ angle. In theory, the $\beta$ should be 54.7° and $\gamma$ should be 0°. The measured result shown in **Figure S6b** confirms that the magnetic field maintained vertical alignment.

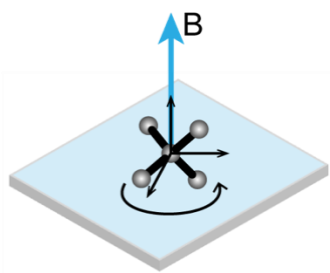
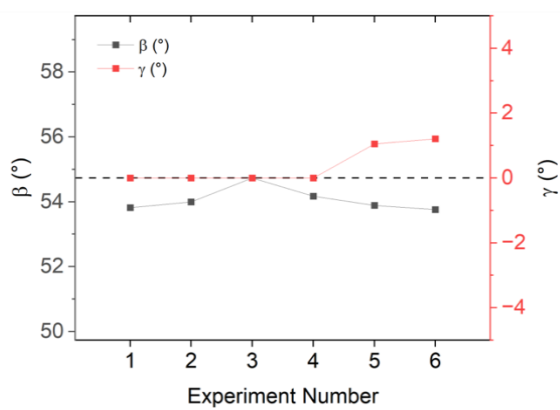

**Figure S6.** Validation of the magnetic field orientation. a) Schematic of the experiment setup. b) Fitting result of $\beta$ and $\gamma$. The horizontal dashed line indicates the theoretical value of $\beta$ and $\gamma$ when the direction of the magnetic field is perpendicular to the sample plane.

**Supporting Information Note 5: Characterizing of samples**

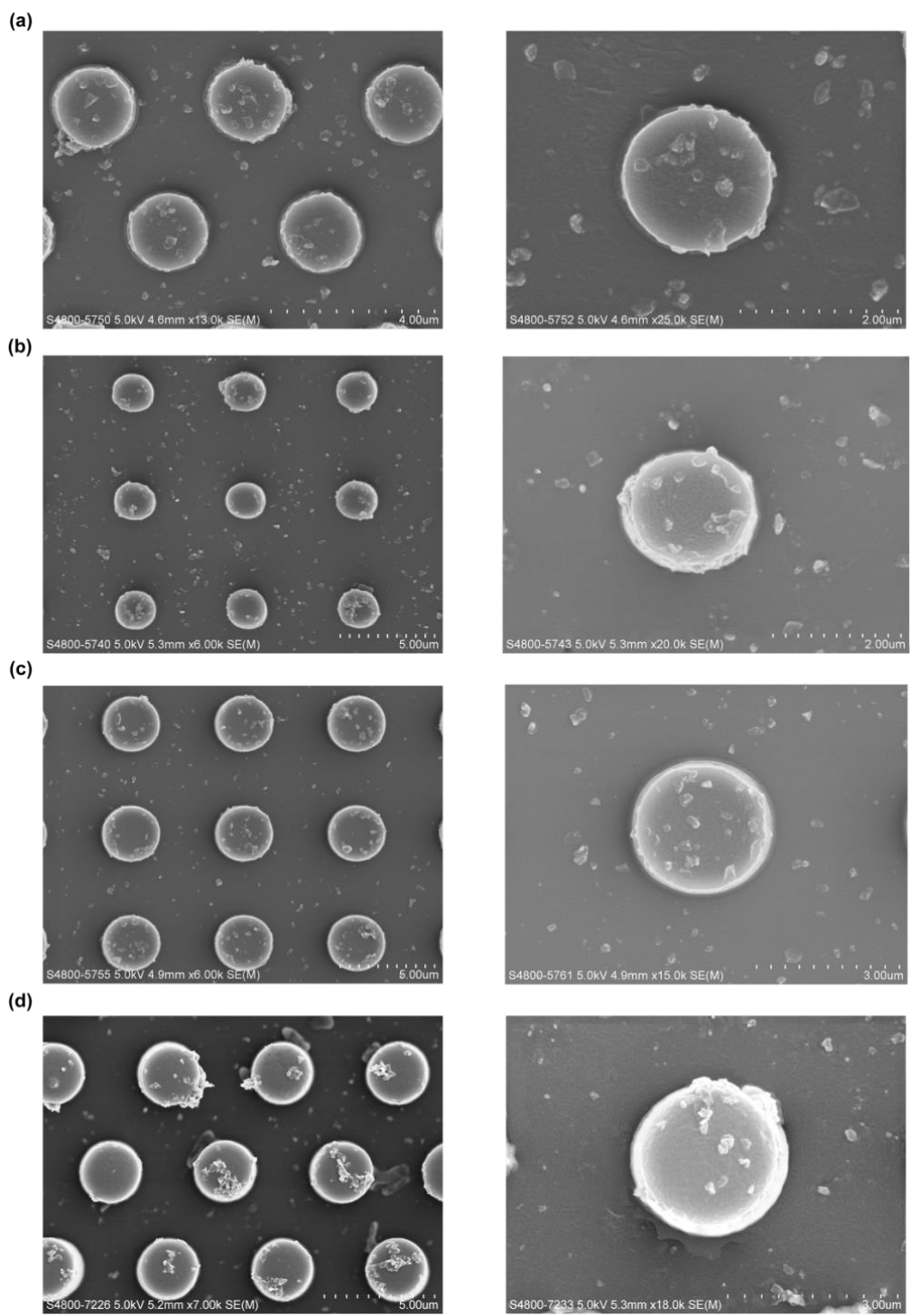

**Figure S7.** SEM images of nanodiamond-coated PDMS pillar arrays with various geometries used in the experiments. All pillars have a height of 6 μm. a) 2 μm diameter, 4 μm center-to-center spacing. b) 2 μm diameter, 6 μm center-to-center spacing. c) 3 μm diameter, 6 μm center-to-center spacing. d) 2.5 μm diameter, 5.5 μm center-to-center spacing. Single-pillar SEM images represent depth-of-field–extended composites formed by merging images acquired at different focal planes (pillar top and substrate) due to insufficient depth of field.

**Supporting Information Note 6: Sensitivity of the rotation angle measurement**

We investigate the angle measurement sensitivity as follows. For the three Euler angles $\alpha, \beta$ and $\gamma$, $\alpha$ is determined from the ODMR-LPM hybrid method, $\beta$ and $\gamma$ are determined from the ODMR method. To estimate the sensitivity, we measured the standard deviation of angles with different integration time. For $\beta$ and $\gamma$ getting from the ODMR method, we fit them with the shot noise function $\sigma = \frac{\eta}{\sqrt{t}}$. Their sensitivities are $0.87\ °\ Hz^{-\frac{1}{2}}$ and $1.76\ °\ Hz^{-\frac{1}{2}}$, respectively. For $\alpha$ getting from the ODMR-LPM hybrid method, the result doesn't show a clear linear relationship between the standard deviation and $t^{-\frac{1}{2}}$, which means that it is not shot noise limited. In fact, due to the measurement method, factors such as the stability of the turntable at different rotation speeds can also influence the measurement error. This may result in shot noise not accounting for the entire contribution to the total noise. The measurement standard deviation is around 1 degree.

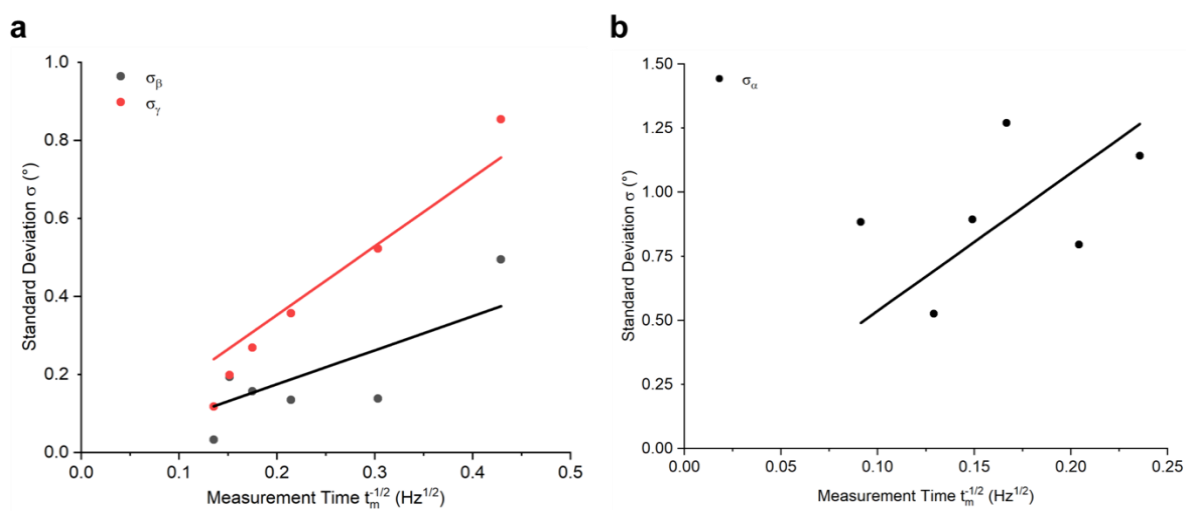

**Figure S8.** Standard deviation (STD) of the Euler angles as functions of the data acquisition time. **a.** STD of $\beta$ and $\gamma$ determined based on the ODMR-LPM hybrid method. The line is the linear fitting with 0 intercept. **b.** The STD of $\gamma$ determined based on the ODMR-LPM hybrid method.

**Supporting Information Note 7: Relocation of the permanent magnet**

During the experiment, the permanent magnet sometimes needs to be moved away and then returned to its original position. Therefore, it was necessary to test the change in the magnetic field direction after relocating the permanent magnet. To investigate this, we selected a single FND and performed ODMR measurements after each magnet relocation. The experiment was repeated 10 times. In principle, only the angles $\beta$ and $\gamma$ obtained from the ODMR method are affected. As shown in **Figure S9**, the change in the measured angles is less than 1°, and the standard deviations of the 10 measurements for $\beta$ and $\gamma$ are 0.10° and 0.34°, respectively. This result indicates that the magnet returns to nearly the same position and that the magnetic field remains almost unchanged after relocation.

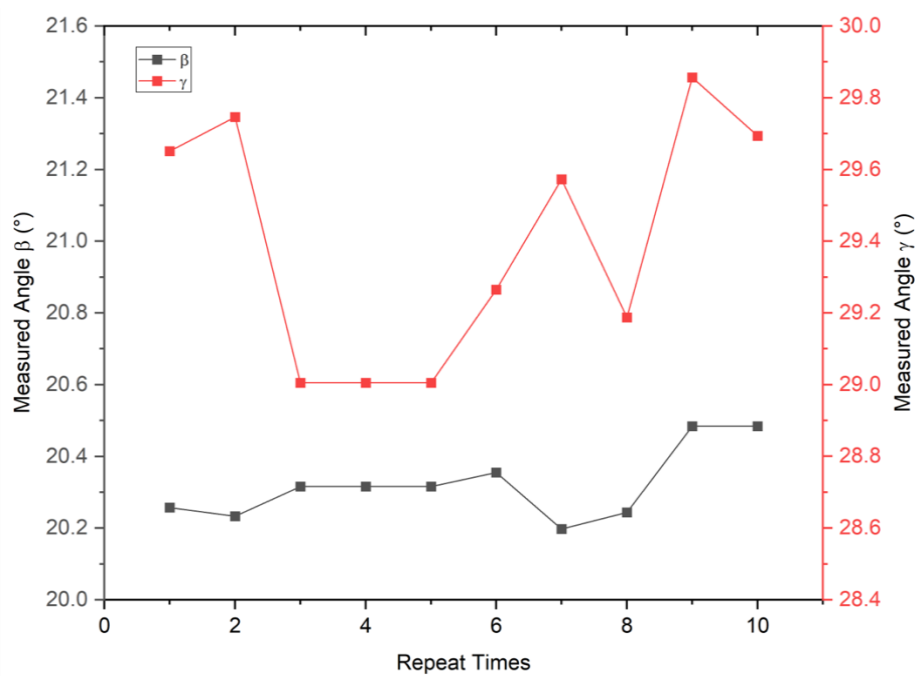

**Figure S9.** Measurement result of $\beta$ and $\gamma$ for the same FND after magnet relocation.

**References**

[1] A. Öchsner, *Classical beam theories of structural mechanics*, Springer, **2021**.
[2] J. M. Gere, S. Timoshenko, Mechanics of Materials. ed, *Boston, MA: PWS* **1997**,
[3] I. Schoen, W. Hu, E. Klotzsch, V. Vogel, Probing cellular traction forces by micropillar arrays: contribution of substrate warping to pillar deflection, *Nano Lett* **2010**, 10, 1823.10.1021/nl100533c
[4] I. Schoen, Substrate-mediated crosstalk between elastic pillars, *Applied Physics Letters* **2010**, 97,
[5] K. Johnson, Contact Mechanics, Cambridge University Press, Cambridge, 1985, **1982**,

[6] M. W. Doherty, N. B. Manson, P. Delaney, F. Jelezko, J. Wrachtrup, L. C. L. Hollenberg, The nitrogen-vacancy colour centre in diamond, *Physics Reports* **2013**, 528, 1.https://doi.org/10.1016/j.physrep.2013.02.001

[7] M. E. Rose, *Elementary Theory of Angular Momentum*, Wiley, **1957**.

[8] M. W. Doherty, J. Michl, F. Dolde, I. Jakobi, P. Neumann, N. B. Manson, J. Wrachtrup, Measuring the defect structure orientation of a single NV centre in diamond, *New Journal of Physics* **2014**, 16, 063067.Artn 063067
10.1088/1367-2630/16/6/063067

[9] L. Rondin, J. P. Tetienne, T. Hingant, J. F. Roch, P. Maletinsky, V. Jacques, Magnetometry with nitrogen-vacancy defects in diamond, *Rep Prog Phys* **2014**, 77, 056503.10.1088/0034-4885/77/5/056503

[10] T. P. M. Alegre, C. Santori, G. Medeiros-Ribeiro, R. G. Beausoleil, Polarization-selective excitation of nitrogen vacancy centers in diamond, *Physical Review B—Condensed Matter and Materials Physics* **2007**, 76, 165205

[11] M. W. Doherty, N. B. Manson, P. Delaney, L. C. L. Hollenberg, The negatively charged nitrogen-vacancy centre in diamond: the electronic solution, *New Journal of Physics* **2011**, 13.Artn 025019
10.1088/1367-2630/13/2/025019

[12] R. J. Epstein, F. M. Mendoza, Y. K. Kato, D. D. Awschalom, Anisotropic interactions of a single spin and dark-spin spectroscopy in diamond, *Nature Physics* **2005**, 1, 94.10.1038/nphys141

[13] V. R. Horowitz, B. J. Alemán, D. J. Christle, A. N. Cleland, D. D. Awschalom, Electron spin resonance of nitrogen-vacancy centers in optically trapped nanodiamonds, *Proceedings of the National Academy of Sciences* **2012**, 109, 13493.doi:10.1073/pnas.1211311109

[14] F. Münzhuber, F. Bayer, V. Marković, J. Brehm, J. Kleinlein, L. W. Molenkamp, T. Kiessling, Polarization-Assisted Vector Magnetometry with No Bias Field Using an Ensemble of Nitrogen-Vacancy Centers in Diamond, *Physical Review Applied* **2020**, 14, 014055.10.1103/PhysRevApplied.14.014055

[15] G. Anzolin, A. Gardelein, M. Jofre, G. Molina-Terriza, M. W. Mitchell, Polarization change induced by a galvanometric optical scanner, *J. Opt. Soc. Am. A* **2010**, 27, 1946.10.1364/JOSAA.27.001946